\documentclass{aa}  
\usepackage{overpic}
\usepackage{graphicx}
\usepackage{txfonts}
\usepackage{lipsum}
\usepackage{subcaption}         
\usepackage{lscape}             
\usepackage{placeins}           

\usepackage[colorlinks=true, linkcolor=blue, citecolor=blue, urlcolor=blue]{hyperref}

\newcommand{\pg}{\citetalias{Perez-Gonzalez2026}}
\defcitealias{Perez-Gonzalez2026}{PG26}

\newcommand{\bb}{\citetalias{Barro2025b}}
\defcitealias{Barro2025b}{B26}

\newcommand{\dg}{\citetalias{deGraaf_review}}
\defcitealias{deGraaf_review}{dG25}

\newcommand{\mm}{\citetalias{Merida2026}}
\defcitealias{Merida2026}{M26}

\begin{document}

   \title{UV-to-optical insights into the BH$^*$ model in little red dots}

\titlerunning{UV-to-optical insights into the BH$^*$ model in little red dots}
%
%
%

      \author{Rosa M. Mérida\inst{1}\corrauth{Rosa.MeridaGonzalez@smu.ca}
        \and Marcin Sawicki\inst{1} \and Chris J. Willott\inst{2} \and Gaia Gaspar\inst{1,3} \and Kartheik G. Iyer\inst{4}
        }

   \institute{  
            Institute for Computational Astrophysics and Department of Astronomy and Physics, Saint Mary's University, 923 Robie Street, Halifax, NS B3H 3C3, Canada \and
            National Research Council of Canada, Herzberg Astronomy \& Astrophysics Research Centre, 5071 West Saanich Road, Victoria, BC, V9E 2E7, Canada \and
            Observatorio Astronómico de Córdoba, Universidad Nacional de Córdoba, Laprida 854, X5000, Córdoba, Argentina \and
            Columbia Astrophysics Laboratory, Columbia University, 550 West 120th Street, New York, NY 10027, USA
            }

   \date{Received September 30, 20XX}

 
  \abstract
  {Little Red Dots (LRDs) are a heterogeneous class of objects, with several proposed scenarios for their physical nature and evolution. While these theories have been tested on individual LRDs using limited spectral features, a systematic Bayesian analysis of the LRD population incorporating the different models across a broad wavelength range is still lacking. In this study, we conduct a consistent ultraviolet (UV)-to-optical continuum fitting analysis of 99 LRDs at $2<z<6$ using \textit{JWST}/NIRSpec PRISM spectroscopy. Employing a modified version of \texttt{Bagpipes}---including blackbody (BB) emission affected by Balmer absorption, stellar and nebular emission attenuated by dust, and an active galactic nucleus (AGN) component---we assess the performance of the black hole star (BH$^*$) model in describing the LRD population. We adopt broad priors and therefore do not impose any specific physical scenario. Our results show that only $\sim4$\% of LRDs with statistically robust solutions (81 objects in total) are best-fit by a BH$^*$ in the optical and a host galaxy in the UV. $\sim9$\% of LRDs show BB-dominated optical continua but lack a stellar component or exhibit AGN UV leakage. Most LRDs are dominated by stellar and/or AGN emission in the optical, with minor BB contribution. When we adopt a prior that disfavors a strong AGN continuum to enforce BH$^*$-like solutions, the percentage of BH$^*$ systems increases to $\sim35$\%, highlighting the strong degeneracy between a BH$^*$ solution and alternative scenarios. Even when BH$^*$-like solutions are enforced, many LRDs still require a stellar-dominated optical continuum. This could indicate limitations in the BH$^*$ model, as further supported by independent evidence, such as the detection of high-ionization emission lines in some LRDs.}


   \maketitle
  \nolinenumbers

\section{Introduction}
\label{sec:intro}

Little Red Dots (LRDs; \citealt{Labbe2023}, \citealt{Barro2024}, \citealt{Greene2024}, \citealt{Matthee2023}) have challenged our understanding of galaxy evolution with their characteristic ``V''-shaped spectral energy distributions (SEDs), compact sizes, absence of X-ray and infrared-to-millimeter emission, and number counts' rapid decline at $z\lesssim2$ (e.g., \citealt{Ananna2024}, \citealt{Casey2024}, \citealt{Kokorev2024}, \citealt{Maiolino2024}, \citealt{Perez-Gonzalez2024}, \citealt{Hviding2025}, \citealt{Kocevski2025}, \citealt{Labbe2025}, \citealt{Perger2024}, \citealt{Setton2025}, \citealt{Ma2025}, \citealt{Zhuang2025}). Since LRDs cannot be straightforwardly explained using traditional combinations of stellar and active galactic nucleus (AGN) emission, new theories have emerged that explore alternative, more exotic physical scenarios.

The black hole star (BH$^*$) model is currently one of the most popular theories (e.g., \citealt{Inayoshi2024}, \citealt{Ji2025}, \citealt{Naidu2025}, see also \citealt{deGraaf2025}, \citealt{Rusakov2025} as supporting observational evidence). This model pictures LRDs as accreting BHs whose emission is reddened by a dense envelope or cocoon of neutral gas (${n_H \sim 10^{9-11} \mathrm{cm}^{-3}}$, $N_H \sim 10^{24-26} \mathrm{cm}^{-2}$), with a near-unity covering factor \citep{Torralba2026a}. Such a gas envelope behaves as a photosphere near the Hayashi limit, with temperatures ${\sim 5,000 - 7,000}$~K (e.g., \citealt{Inayoshi2024}, \citealt{Kido2025}, \citealt{Inayoshic}).

From the observational perspective, this model is a good representation of the optical continuum of LRDs (see \citealt{deGraaf_review,deGraaf2025}, \citealt{Ji2025}, \citealt{RLin2025}, \citealt{Taylor2025}, \citealt{Barro2025b}, \citealt{Merida2026}, \citealt{Perez-Gonzalez2026}, \citealt{Sun2026}, \citealt{Umeda2026}, among many others). However, the nature and broadening of the Balmer lines in LRDs is still a matter of debate (e.g., electron scattering through nuclear dense plasma, \citealt{Rusakov2025}; broad line region BLR stratification, \citealt{Scholtz2026}; hard, ionizing-rich intrinsic SEDs and an anisotropic radiation field, \citealt{Madau2026}; young massive stars surrounding the envelope, \citealt{Asada2026}). Consequently, it is still unclear whether the BH$^*$ model provides the means to explain the nebular emission of LRDs.

The rest-frame ultraviolet (UV) part of the ``V'' shape is also yet to be explored in more detail in the context of the BH$^*$ model.
This emission is proposed to be powered by a low-mass, metal and dust-free host, with stellar masses $M_\star\sim10^{7-8}\,M_\odot$ (e.g., \citealt{Matthee2025}, \citealt{Pizzati2025}, \citealt{Lin2026} via clustering analysis; see also \citealt{Korber2026}, \citealt{Sun2026}, although see \citealt{Barro2025b} reporting $M_\star>10^9\,M_\odot$ and $A_V\sim0.5$~mag for some LRDs). However, a full and consistent UV-to-optical spectral fitting of a sample of LRDs is still required to achieve a more complete understanding of LRDs' hosts and their contribution to the ``V'' shape.

\citet{deGraaf_review} (\dg\, hereafter), \citet{Barro2025b} (\bb\, hereafter), and \citet{Perez-Gonzalez2026} (\pg\, hereafter) are pioneer examples providing statistical analyses of LRDs based on large spectroscopic samples. \dg\, modeled the LRD optical continuum using a modified blackbody (BB) prescription (see also \citealt{Umeda2026}), but did not model the UV emission of their sample. 

\bb\, and \pg\, fitted the LRD UV‑to‑optical flux following a semi‑empirical approach, assuming a stellar origin for the UV continuum and a \textit{Cliff}-like nature for the optical. 
\textit{The Cliff} \citep{deGraaf2025} is an LRD at $z_{\mathrm{spec}}=3.55$ that exhibits an extreme break at the Balmer limit (Balmer strength $\sim6.9$), which, given its compact size, cannot be attributed to a combination of stellar and AGN emission. In short, \bb\, fitted the optical continuum of \textit{The Cliff} with a low-order polynomial and applied it, together with a dust-attenuation component, to reproduce the optical continuum of the LRDs in their sample. \pg\, also used an empirical LRD-core template based on the LRD exhibiting the most pronounced ``V''-shaped continuum in their sample.

In \citet{Merida2026} (\mm\, hereafter), we improved upon previous attempts at UV-to-optical spectral fitting. We used a modified version of the \texttt{Bagpipes} spectral‑fitting code \citep{Carnall2018} to obtain a consistent, data-driven description of the spectrum of \textit{The Cliff}, as observed with the Near Infrared Spectrograph (NIRSpec; \citealt{Jakobsen2022}), onboard the \textit{James Webb Space Telescope} (JWST; \citealt{Gardner2023}), in PRISM mode (R~$\sim$~100). This modified version of \texttt{Bagpipes} allows a BB spectrum component to be fitted, with a peak temperature ranging from 1,000 to 7,000~K. In our methodology, this BB component is subject to gas absorption near the Balmer limit. 
An essential aspect of our approach is that neither a BB-dominated optical continuum nor a stellar origin for the UV emission is imposed a priori. Instead, we let the code select the contribution of each model component, including also AGN emission. This treatment of the envelope is thus more flexible than a semi-empirical approach.

A central outcome of this method is the suppressed AGN continuum in \textit{The Cliff}, which---as noted earlier---is not enforced by the priors but was naturally recovered by the fit.
The BH$^*$ model provides a statistically robust solution for this LRD, yielding a BB temperature $\sim4,500$~K and an optical continuum dominated by the BB spectrum. The UV is powered by the host galaxy, spatially unresolved in JWST imaging but recovered through the fit. 
However, some level of dust attenuation, acting on the stellar and nebular components, is required to fit the UV continuum emission of this LRD ($A_V\sim0.5-1$ mag depending on the selected dust attenuation law), ruling out a dust-free scenario for \textit{The Cliff}.

Another key finding of the \mm\, analysis of \textit{The Cliff} is that the solutions are degenerate in the UV. For this LRD, a purely stellar UV continuum and a mixed stellar + AGN origin are statistically indistinguishable options. This is essential since AGN UV leakage is not expected within the BH$^*$ framework, as the envelope is assumed to reprocess all of the AGN continuum. However, the leakage scenario cannot be fully ruled out given the detection in some LRDs of hard‑ionization UV lines (\citealt{Labbe2024environment}, \citealt{Tang2025}, \citealt{Treiber2025}) and Fe\,II emission, typically associated with classical AGNs (e.g., \citealt{Labbe2024environment}, \citealt{Tripodi2025}, \pg, \citealt{Torralba2026}). Additionally, \citet{Sok2026} studied the narrow-line emission of $\sim20$ LRDs, finding that at least 40\% of them show ratios compatible with a high ionization parameter and electron temperature. These findings open the possibility that the BH$^*$ envelope may not be perfectly spherical or may exhibit a patchy structure (e.g., \citealt{Hviding2026}, \citealt{Sok2026}, \citealt{Sneppen2026b}). Such geometries would allow for partially transparent lines of sight through which these emission lines, and perhaps a small fraction of the AGN continuum, could escape.

However, it remains to be seen whether these results for \textit{The Cliff}, pointing to moderate dust levels in the host and possible AGN UV leakage, are representative of the full LRD population.
The availability of spectroscopic data from NIRSpec using PRISM for hundreds of LRDs (with medium or high resolution additionally available for half of them) enables, for the first time, a statistical characterization of the LRD population, as well as consistent UV-to-optical spectral fitting analyses. 

In this work, we follow the method presented in \mm\, and use PRISM spectra to probe the rest-frame UV and optical continuum of LRDs at $2<z<6$. We examine the relative contributions of the AGN, stellar, and BB components to the total LRD emission, assessing whether any tensions arise for the BH$^*$ scenario or, alternatively, whether this model provides a physically consistent description for a wide range of LRDs.
In Sect.~\ref{sec:sample}, we describe the LRD selection process and compare our sample with the datasets from \dg\, and \bb. Section~\ref{sec:method} summarizes the method implemented in this work, based on \mm.
Section~\ref{sec:results} presents the diverse demographics that emerge when no specific physical scenario is imposed during the fitting, and describes the resulting properties of our LRD sample. Section~\ref{sec:discussion} evaluates the viability of the best-fitting solutions and the trends reported in Sect.~\ref{sec:results}, and also examines the fits and inferred properties obtained when a BH$^*$ solution is enforced through the priors. Finally, Sect.~\ref{sec:conclusions} presents the summary and main conclusions of this work.
Throughout this work, we assumed $\Omega_\mathrm{M,0}=0.3$, $\Omega_{\Lambda,0}=0.7$, and H$_0=70$ km s$^{-1}$ Mpc$^{-1}$. All $M_\star$ and star formation rate (SFR) estimates assumed a \citet{Chabrier2003} initial mass function.

\section{Sample selection}
\label{sec:sample}

This work is based on JWST NIRSpec/PRISM data, which provides $0.6-5.3\,\mu\mathrm{m}$ spectral coverage, using the micro-shutter assembly. Data were obtained from version 4 of the DAWN JWST Archive (DJA; \citealt{Brammer2025}). We selected \texttt{grade=3} sources, i.e., the highest spectral quality flag based on visual inspection, corresponding to high-quality data and robust redshifts, and imposed no signal-to-noise (S/N) cut. This led to a parent sample of 37,290 NIRSpec spectra at ${0.1 < z < 11}$. 

Once only PRISM data were selected, the sample was screened to include only LRD candidates following \dg. In short, we computed the rest-frame UV and optical slopes ($\beta_{\mathrm{UV}}$ and $\beta_{\mathrm{opt}}$) and applied the following cuts to ensure the characteristic ``V''-shaped spectra of LRDs (see also \citealt{Hviding2025}):

$$\beta_{\mathrm{opt}} > 0;\; \beta_{\mathrm{UV}} < -0.2;\, \beta_{\mathrm{opt}} - \beta_{\mathrm{UV}} > 0.5$$

As mentioned in \dg, the pivot point of the ``V'' shape, used to measure the slopes and located around the Balmer limit (see \citealt{Setton2024}), varies from galaxy to galaxy. For each source, we identified this pivot wavelength by searching for an inflection point around the Balmer limit.

After masking the emission lines, the continuum was smoothed using a median filter (\texttt{medfilt} from \texttt{scipy}, \citealt{Virtanen2020}) with an $11\times11$ pixel kernel, reducing the noise while preserving the underlying continuum shape. We then computed the local slopes in log$-$log space. The pivot point, selected within the $3,000-5,000~\AA$ wavelength range, was defined as the wavelength where the slope changes from negative to positive and remains positive for at least five consecutive points. This transition is taken as the ``V'' shape's pivot point and is used to separate the rest‑frame UV and optical regimes. If no such transition was found, which can happen when the noise level exceeds what can be mitigated by the smoothing, the pivot point was instead selected as the wavelength where the slope reaches its most negative value, corresponding to the point of strongest curvature.

Once the pivot point was identified, the UV and optical slopes were measured through weighted linear fits in log$-$log space, i.e., log (f$_\lambda)$ vs log ($\lambda$/$\lambda_\mathrm{pivot}$), using the flux uncertainties as weights in the fit. A minimum of 20 data points was required in each part of the ``V'' shape. The UV window was defined between $1,300~\AA$ and the pivot point, while the optical window was defined between the pivot point and $7,500~\AA$. The resulting slopes were then used to evaluate the LRD selection criteria.

Additionally, we only considered objects with broad H$\alpha$ emission (or H$\beta$ when H$\alpha$ is not included in the NIRSpec spectra). For this purpose, we used a method that provides a fast and efficient first-pass screening, as we aim to identify broad-line candidates rather than to perform detailed modeling of individual line profiles. For each line, we considered a 70~$\AA$ rest-frame window centered on the lines, computed the local continuum and subtracted it from the spectrum. We then identified the line peak and determined the full width at half maximum (FWHM) by linearly interpolating the wavelengths at which the continuum-subtracted flux crosses half the peak value. The FWHM was converted to velocity, and a 1,000 km/s cut was imposed. In addition to this cut, we also imposed a S/N~$>3$ at the line peak.
These screenings left us with 224 LRD candidates. We then removed duplicates, as some galaxies were observed in more than one survey and are linked to more than one identifier. This yielded 207 unique LRD candidates. 

Next, we applied a compactness cut based on the sources' emission in the $F444W$ NIRCam band (see \citealt{Kokorev2024}, \citealt{Kocevski2024}). We defined compactness as the ratio of the flux measured in 0.5 and 0.3\arcsec\, apertures and imposed compactness~$<1.7$ \citep{Merida2025}. For this purpose, we used JWST/NIRCam image mosaics from the DJA (version 7; pixel scale of 0.04\arcsec) when available; otherwise, we relied on the image repositories of the surveys listed below. We performed background subtraction using \texttt{photutils} \citep{Bradley2016} when background-subtracted images were not provided. 

These mosaics were obtained as part of the CAnadian NIRISS Unbiased Cluster Survey (CANUCS; GTO-1208, PI: Willott; \citealt{Willott2022}); the Public Release IMaging for Extragalactic Research (PRIMER; GO-1837; PI: Dunlop); the Ultradeep NIRSpec and NIRCam ObserVations before the Epoch of Reionization (UNCOVER; GO-2561; PIs: Labbe \& Bezanson; \citealt{Bezanson2024}); the North ecliptic pole EXtragalactic Unified Survey (NEXUS; G0-5105; PI: Shen; \citealt{Shen2024}); the Parallel wide-Area Nircam Observations to Reveal And Measure the Invisible Cosmos (PANORAMIC; GO-2514; PI: Williams; \citealt{Williams2025}); the JWST Advanced Deep Extragalactic Survey (JADES; GTO-1180, 1181, 1210, 1286, GO-1895, 1963; \citealt{Eisenstein2026}); and the Cosmic Evolution Early Release Science (CEERS; ERS-1345; PI: Finkelstein; \citealt{Finkelstein2025}).
We note that $F444W$ NIRCam coverage was not available for 25/207 of these candidates.
The final screened sample consisted of 99 LRDs, located at ${2<z<6}$. Six LRDs at $z>6$ were dismissed, as we require enough optical continuum coverage to properly constrain the BB emission (see also \dg). 

\begin{figure}[htp]
\centering
\begin{overpic}[width=\columnwidth]{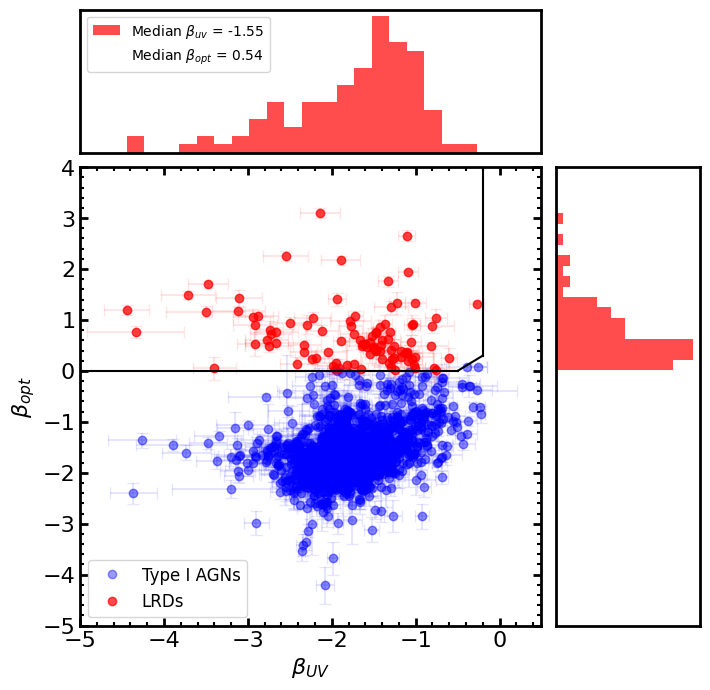}
\setlength{\fboxrule}{0.8pt}   
\setlength{\fboxsep}{0pt}      
    \put(63,10){\fbox{\includegraphics[width=0.35\columnwidth]{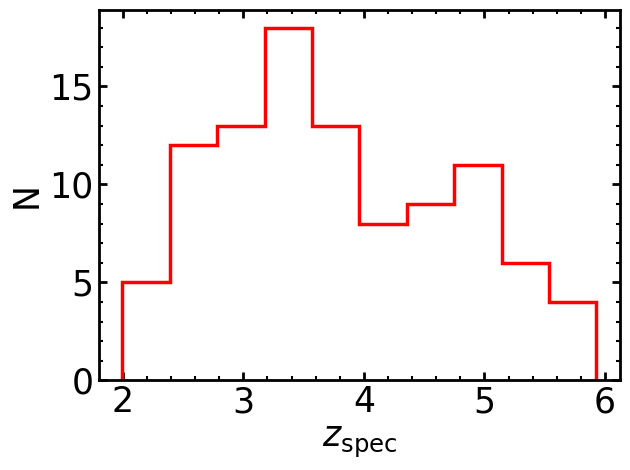}}}
\end{overpic}
\caption{$\beta_{\mathrm{UV}}$ vs $\beta_{\mathrm{opt}}$ of our sample (red), consisting of 99 LRDs. Compact non-LRD Type I AGN extracted from the parent sample are shown in blue. The demarcation line between the LRD and non-LRD loci is highlighted in black. The redshift distribution of the LRDs is included as an inset.}
\label{fig:z_slopes}
\end{figure}

Figure~\ref{fig:z_slopes} shows the spectroscopic redshifts of our sample and its location in the $\beta_{\mathrm{UV}}$ vs $\beta_{\mathrm{opt}}$ plane, together with a dataset of compact Type 1 AGN extracted from the parent sample. These Type I AGNs are non-LRD objects at $2<z<6$ that show broad Balmer emission but lack a ``V''-shaped spectrum (1,381 objects in total).

26 of the LRDs in our dataset belong to the Red Unknowns: Bright Infrared Extragalactic Survey (RUBIES; GO-4233; PIs: de Graaff \& Brammer; \citealt{deGraaf2025Rubies}); 21 were observed as part of JADES \citep{DEugenio2024}; 12 are from the CANDELS-Area Prism Epoch of Reionization Survey (CAPERS; GO-6368; PI: Dickinson); 11 are from NEXUS; 8 are from UNCOVER; 7 are from CANUCS; 7 are from NIRSpec GTO-Wide (GTO-1211; PI: Isaak; \citealt{Maseda2024}); 2 belong to Mirage or Miracle? (MoM; GO-5224, PIs: Oesch \& Naidu); 2 are from Quiescent or dusty? Unveiling the nature of extremely red galaxies at $z>3$ (GO-2198; PI: Barrufet \citealt{Barrufet2025}); 1 is from The High-$z$ Menagerie: A Rare Chance to Study the Early and Exotic Transient Universe (DD-6585; PI:Coulter; \citealt{Coulter2024}); 1 is from the Spectroscopic follow-up of ultra-high-$z$ candidates in CEERS (DD-2750; PI: Arrabal Haro; \citealt{Arrabal2023}); and 1 belongs to Extremely massive galaxies in the early universe: a challenge to Lambda-CDM (GO-4106; PI: Nelson). 
For galaxies in cluster fields (i.e., objects from CANUCS and UNCOVER), the physical properties that we report in this work were corrected for lensing magnification using the best-fit lensing magnification factors reported in the CANUCS DR1 \citep{Sarrouh2025} and the UNCOVER DR3 SUPER \citep{Suess2024} catalogs.

\subsection{Cross-matching with other spectroscopic LRD samples}
\label{sec:match_samples}

We matched the 99 LRDs selected earlier in Sect.~\ref{sec:sample} with the LRD sample from \dg\, and \bb, which provide a list of identifiers for their respective objects. 49 out of the 99 LRDs at ${2<z<6}$ from \dg\, are present in our dataset. From the remaining objects, we could not find $F444W$ NIRCam coverage for 5 of them. This screening does not account for the associated duplicates, which are not removed from the total 99 objects. Additionally, 8 of \dg\, LRDs are not present in our parent sample. These are objects from MoM that are not publicly available in DJA. 

Besides the duplicates or the lack of data, some of the \dg\, LRDs did not pass our screenings because (a) we did not measure a clear broad component (i.e., line velocities $>$ 1,000 km/s) in their Balmer lines based on PRISM using our method (e.g., DJA \texttt{srcid} = 4111589, 2015) or (b) we measured a $\beta_{\mathrm{opt}} < 0$. This last possibility takes place when $\beta_{\mathrm{opt}}$ is close to 0 (e.g., \texttt{srcid} = 149501, 61496). In those cases, the definition of the inflection point can lead to slight variations in classification. 

Additionally, the definition of compactness in \dg\, differs from ours, since they used 0.1 and 0.2\arcsec\, apertures. They also performed a two-component point source and S\'ersic profile fitting when possible. Consequently, objects close to the compactness limit according to our criterion may not be selected following their approach. However, the aim of this work is also to understand how the physical properties of LRDs evolve with compactness. Therefore, it is important to include objects with potential extended emission that meet the average compactness criteria, since they may be included in more conservative LRD selections.

The \bb\, sample contains 77 LRDs at $2 < z <6$, of which 46 are in common with \dg, and 38 are in common with our sample. 14 LRDs from our sample are only present in \dg, whereas 3 are only present in \bb. This sums up a total of 52 LRDs in common with the set of \dg\, and \bb\, galaxies (\dg$\cup$\bb). A comparison of the properties retrieved in those studies and the values reported in this work is included in Appendix A\footnote{\href{https://doi.org/10.5281/zenodo.22057251}{The complete appendices are available on Zenodo}}.

39/77 LRDs from \bb\, did not pass our slope and broad-line component screenings. These correspond to cases where $\beta_{\mathrm{opt}}\sim0$ (e.g., \texttt{srcid} = 2783, 3284) and a lack of a broad component as measured from PRISM data (e.g., \texttt{srcid} = 30148719). We found NIRCam coverage for the 38 remaining objects.
Note that \bb\, imposed ${f(0.5\arcsec)/f(0.2\arcsec) < 1.5}$, with $f$ being the flux in the $F444W$ NIRCam band, which is a less conservative compactness limit than ours. 

\begin{figure}
    \centering
    \includegraphics[width=0.9\linewidth]{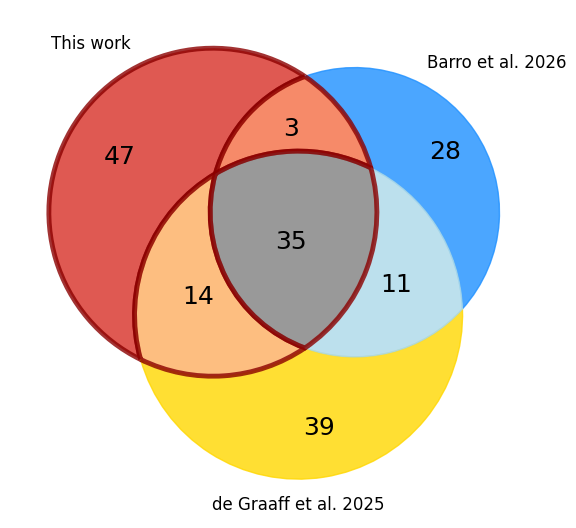}
    \caption{Venn diagram illustrating the intersections between our LRD sample (red) and the LRD samples from \dg\, (yellow), and \bb\, (blue). The figure quantifies both the shared objects and the subsets unique to each catalog.}
    \label{fig:venn}
\end{figure}

Fig.~\ref{fig:venn} shows a Venn diagram illustrating the overlap between our LRD sample and the samples from \dg\, and \bb. These comparisons highlight how heterogeneous LRD selections can be, even when similar criteria are applied. Nevertheless, our final sample provides a reliable and representative census of objects broadly considered LRDs.

\section{Methodology}
\label{sec:method}

In order to derive the physical properties of our LRD sample, we followed the spectral fitting method presented in \mm, which is based on a modified version of \texttt{Bagpipes}. To include the ingredients of the BH$^*$ model, our custom version of \texttt{Bagpipes} incorporates some built-in functions and additional custom components. 

The model components work as follows. For the stellar model, we selected non-parametric star formation histories (SFHs) through the \texttt{continuity} mode \citep{Leja2019}.
For the dust component, we assumed a Small Magellanic Cloud (SMC; \citealt{Gordon2003}) dust attenuation law (see \mm\, for a comparison between SMC and a \citealt{Calzetti2000} law in the context of \textit{The Cliff}). 

We used a prior configuration that potentially allows massive, metal-rich, and dusty solutions for the host galaxy. In \mm, we referred to this setup as the BH$^*$+dusty host configuration, where it was tested in the context of \textit{The Cliff}. There, we showed that a dust-free host solution cannot properly fit the UV continuum emission of this object. Therefore, in the present work, we let the code explore a broad range of $A_V$ values. 

The AGN is modeled as a broken
power law, with $\alpha_\lambda$ and $\beta_\lambda$ controlling the UV and optical slopes, \texttt{f5100A} setting the flux at 5,100$\AA$, and \texttt{hanorm} controlling the intensity of the Balmer lines. The width of the lines was fixed through the H$\alpha$ line width as measured by us from the spectra. 

The \texttt{f5100A} parameter can act to suppress or enhance the AGN continuum emission. This parameter was left free in the code, following \mm. We refer to this approach as ``AGN-agnostic''. In Sect.~\ref{sec:discussion_2} we force the BH$^*$-model by manually suppressing the AGN continuum through this \texttt{f5100A} parameter (referred to as ``AGN-suppressed'' in Sect.~\ref{sec:discussion_2}).

\begin{table*}[htp]
\setlength{\tabcolsep}{1.6pt}  
\tiny
    \centering
    \caption{\texttt{Bagpipes} priors used in this work}
    \begin{tabular}{c|c|c|c|c|c}
    \hline
         SFH&Nebular&Dust&AGN&BB&Absorption\\ \hline\hline
         \texttt{bin$\_$edges} = [0, 10, 30, 100, 300, 700]&\texttt{log U} = [$-4, 0$]& $A_V$ = [0, 5]&\texttt{alphalam} = [$-2.8, \,2$]&\texttt{T} = [1000, 7000]&\texttt{log sigma$\_$B0} = [$-18, -15$]\\
         \texttt{massformed} = [4, 10]&&&\texttt{betalam} = [0.8, 1.6]&\texttt{log A} = [$-25, -16$]&\texttt{log N} = [19, 24]\\
         \texttt{metallicity} = [0.001, 2.5]&&&\texttt{hanorm} = $[0, \,5\times10^{-16}]$&&$\lambda_{break}$ = [3000, 4000]\\   
         &&&\texttt{f5100A} = [0, 3$\times10^{-16}$]&&\texttt{break$\_$width} = [10, 100]\\ \hline
          
    \end{tabular}
    \label{tab:priors}
    \tablefoot{The SFH is set to the \texttt{continuity} mode, and the bins are expressed in Myr. dSFR$_i$, which represents the change in log SFR between two consecutive SFH bins, is set to $[-10, 10]$, following a Student-t distribution. The \texttt{massformed} parameter is logarithmic and in $M_\odot$ units; the metallicity is logarithmic and in $Z_\odot$ units; $A_V$ is in magnitudes; \texttt{hanorm} has units of erg/s/cm$^2$ and \texttt{f5100A} is in erg/s/cm$^2/\AA$; $\lambda_{break}$ and \texttt{break$\_$width} are in units of $\AA$. The normalization factor of the BB, A, is expressed in erg/s/cm$^2/\AA$ and the temperature is in K. In the absorption component, the \texttt{sigma$\_$B0} $\times$ \texttt{N} product is dimensionless. In total, we worked with 18 free parameters.}
\end{table*}

We explored a wide range of UV slopes ($\alpha_\lambda$), which naturally allows AGN UV leakage even when the code selects a relatively low \texttt{f5100A} normalization in the AGN-agnostic setup. By contrast, in the AGN‑suppressed configuration, such leakage cannot occur because the imposed normalization is forced to be extremely low.

We then included additional model components using the option \texttt{extra$\_$components}. These additional components correspond to a BB spectrum and a gas absorption feature that acts on the BB at the Balmer limit. 
We modeled the emission from the BB analytically using Planck's law, introducing two parameters: temperature and a normalization factor. We included a gas absorption component to account for the sharp drop in the continuum caused by strong breaks at the Balmer limit. This absorption is controlled by a column density $N$, a cross-section $\sigma_B$, the inflection point wavelength $\lambda_{break}$, and a width parameter that adjusts how sharp or smooth the absorption is. This absorption feature was modeled similarly to the Lyman break, assuming a $\sigma_B\propto (\lambda / \lambda_{break})^3$. A detailed description of the formulation adopted for this component is provided in \mm.

In \mm, we derived two fits per galaxy: one based on the full spectrum and one based only on the continuum. We did this to isolate the effect of the emission lines, whose origin is still under debate (see Sect.~\ref{sec:intro}). The \texttt{Bagpipes} spectral fitting analysis presented in this work is based on fits to the continuum only. As a result, even though Balmer lines are included in the AGN model, they have no impact on the resulting best‑fitting solution. We leave the spectral fitting of continuum + emission lines for future work, once a better understanding of the lines' origin and the mechanisms involved is achieved. We briefly discuss the [OIII] line emission and Balmer break strengths of the LRDs in our sample in Sect.~\ref{sec:elines}.

As a result, we work with 18 free parameters. A summary of the prior configuration is provided in Table~\ref{tab:priors}, and further details on this method can be consulted in \mm.

\section{Results}
\label{sec:results}

\subsection{Demographics}
\label{sec:demographics}

The dataset presented in this work is far from being complete. It is based on available public PRISM and $F444W$ NIRCam data and thus is biased toward bright or peculiar LRDs that were selected for follow-up with NIRSpec. As a result, our interpretations may be representative of the most extreme LRD population, while the lower-luminosity regime remains largely unexplored and may exhibit different physical properties from those inferred for their brighter counterparts. Moreover, slightly different definitions of the inflection point or the compactness criterion can yield different sample selections (see Sect.~\ref{sec:match_samples}). 

\begin{figure*}[htp]
    \centering
    \includegraphics[width=\linewidth]{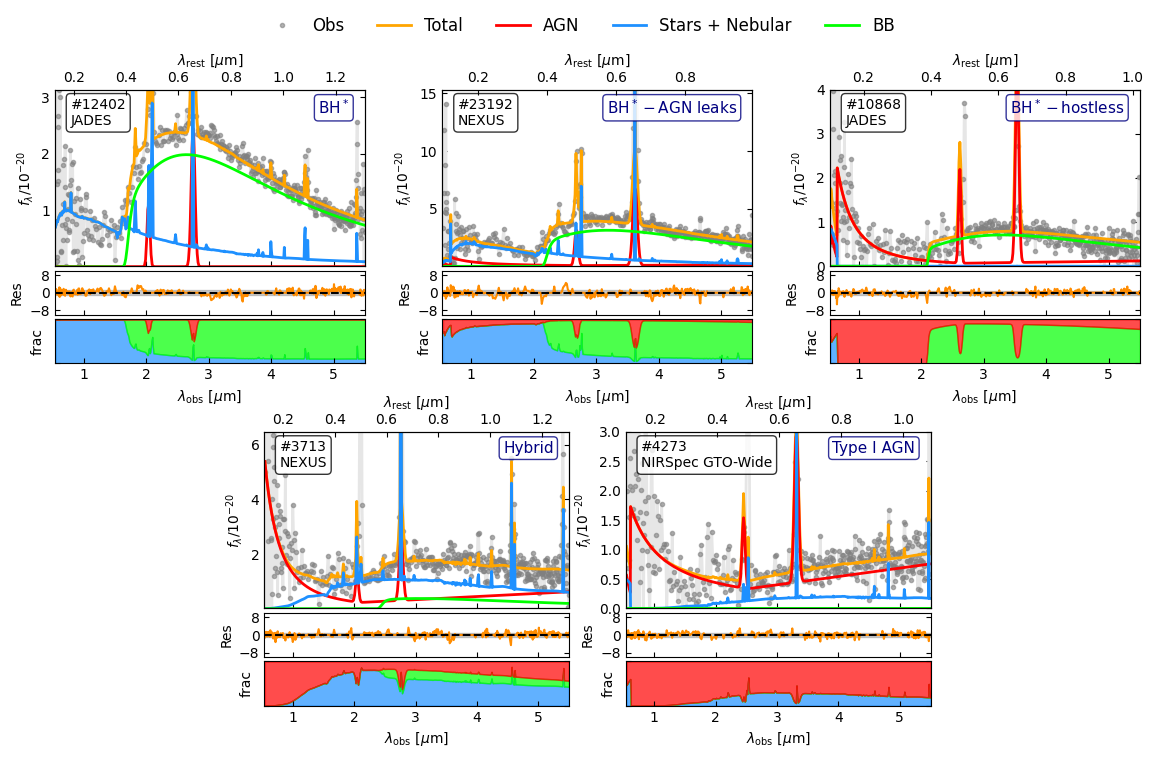}
    \caption{Best-fitting models from \texttt{Bagpipes} (``AGN-agnostic'' approach) based on the continuum emission of a subset of 5 LRDs, that are illustrative of the BH$^*$, BH$^*$-AGN leaks, BH$^*$-hostless, Hybrid, and Type I AGN categories (see Table~\ref{tab:classes} and Sect.~\ref{sec:demographics}). Fluxes in all panels are expressed in units of erg/s/cm$^2/\AA$. In each panel, the top row shows the observed spectrum (gray), the total best-fitting model (orange), the AGN component (red), the stellar + nebular component, attenuated by dust (blue), and the composite of the BB and Balmer absorption (green). The associated DJA identifiers and surveys are included at the top. The residuals of the fit are plotted in the second row. The shaded gray region depicts the standard deviation. The fractional contribution of each component to the total best-fitting model is displayed in the last row, following the same color code as above. In the ``AGN-agnostic'' setup, we get many different combinations of AGN + stars + BB emission, most of them not meeting the BH$^*$ model requirements.}
    \label{fig:example_classes}
\end{figure*}

At the same time, this dataset is intrinsically highly heterogeneous. While some LRDs, such as \textit{The Cliff}, exhibit extreme Balmer breaks, others display less pronounced ``V''-shaped SEDs (e.g., Fig.~\ref{fig:example_classes}, see also \pg). Some show strong collisional lines, such as [OIII], while others only exhibit hydrogen lines, with a negligible contribution from other optical lines (see Sect.~\ref{sec:elines}). 
It is possible that no single model is capable of reproducing the full variety of LRDs. However, the BH$^*$ model is hypothesized to provide a universal framework for understanding all LRDs, and that is the focus of our investigation in this work.

As a result of our fitting, following the method described in Sect.~\ref{sec:method} (``AGN-agnostic''), we were able to recover statistically robust fits for 81/99 LRDs. The remaining 18 objects show large reduced $\chi^2$ values ($\chi_{\mathrm{red}}^2>3$).
Within these 81 LRDs, we were able to identify different subtypes based on how well the BH$^*$ scenario was represented by the best-fitting solution. Note that we are not imposing the BH$^*$ model, since we assumed broad priors for all the parameters.  

In general, most of the fits did not exactly match the expectations from the BH$^*$ model, with an important AGN contribution in some cases, a lack of host emission, or a lack of BB flux in the fit. To better understand these differences and delve deeper into the physical drivers of the LRD continuum emission, we classified these 81 objects into 5 classes, summarized in Table~\ref{tab:classes}.

\begin{table}[htp]
\renewcommand{\arraystretch}{1.2}
    \setlength{\tabcolsep}{1.2pt} 
\footnotesize
    \centering
    \caption{LRD classification based on the AGN, stellar, and BB contribution to the \texttt{Bagpipes} best-fitting model.}
    \begin{tabular}{c|c|c|c|c|c|c|c}
    \hline
    &\# Objects&\multicolumn{2}{|c|}{AGN}&\multicolumn{2}{|c|}{Stars}&\multicolumn{2}{|c}{BB}\\
    \hline
        &$\chi^2_{\mathrm{red}}<3$&UV&Opt&UV&Opt&UV&Opt\\  \hline\hline
        BH$^*$&3&$<10\%$&$<10\%$&$>90\%$&$<20\%$&&$>70\%$\\ \hline
        BH$^*$-AGN leaks&3&$<50\%$&$<10\%$&$>40\%$&$<20\%$&&$>70\%$\\ \hline
        BH$^*$-hostless&4&$>70\%$&$<30\%$&$<10\%$&$<10\%$&&$>70\%$\\ \hline
        Type I AGN&16&&&&&&$<10\%$\\ \hline\hline
        Hybrid&55&&&&&&\\ \hline
    \end{tabular}
    \tablefoot{We divide the contribution of each component into rest-frame UV and optical. The BB does not contribute to the UV by construction due to the gas absorption component. The second and third classes, BH$^*$-AGN leaks and -hostless, correspond to quasi-BH$^*$ configurations in which the optical continuum is mainly powered by the BB, but the UV is partially or totally driven by the AGN continuum. The fourth class corresponds to galaxies with a negligible contribution from the BB. Objects that do not match any of these criteria were classified as Hybrid (fifth class), in between a BH$^*$-like LRD and a compact Type I AGN.}
    \label{tab:classes}
\end{table}

Within the theoretical BH$^*$ framework, the optical continuum is expected to be powered by the BB component, while the UV emission is attributed to the stars (although see \citealt{Zhang2025} or \citealt{Sun2026}, reporting a contribution of the BH$^*$ component to the UV emission). These assumptions define our first class, \textit{BH$^*$}. 
To explore the possibility of AGN UV leakage, which is not allowed within the strict BH$^*$ scenario, we reduced the contribution of the stars to the UV criterion and instead permitted some AGN flux. These criteria define the quasi‑BH$^*$ class \textit{BH$^*$-AGN leaks}.

We then extended this idea to investigate LRDs with a negligible host contribution while still exhibiting a BH$^*$-like optical continuum powered by the envelope. To this end, we selected LRDs that show a strongly suppressed stellar component in the fit, meaning that the UV emission is primarily driven by the AGN. These criteria characterize the quasi‑BH$^*$ class \textit{BH$^*$-hostless}.

We also selected objects in which the BB component is negligible in the fit. These systems correspond to compact \textit{Type I AGN} LRDs, consistent with being Little Blue Dots (e.g., \citealt{Asada2026}, \citealt{Brazzini2026}, \citealt{Madau2026}, \citealt{Scholtz2026}). 

LRDs showing some BB contribution in the optical (i.e., $>10\%$) but not meeting the criteria of the previous classes were grouped into the \textit{Hybrid} category. These objects occupy an intermediate regime between a BH$^*$-like LRD and a compact Type I AGN. 

In total---considering only fits with $\chi_{\mathrm{red}}^2<3$---we identified 3 pure BH$^*$ systems, 3 BH$^*$ with AGN UV leakage, 4 BH$^*$-hostless LRDs, 16 Type I AGNs, and 55 Hybrid objects. 
Consequently, only a small fraction of the sample naturally matches the BH$^*$ picture, while most of the sources fall between a BH$^*$ and a Type I AGN description. Figure~\ref{fig:example_classes} presents one example of each of these categories. The properties of these five representative LRDs, all of which are also members of the \dg$\cup$\bb\, sample, are listed in Table B.1 (``AGN-agnostic'' column) in Appendix B. The fits for the rest of the objects, divided in categories, can be consulted in Appendix C.

\begin{figure*}[htp]
    \centering
    \includegraphics[width=\linewidth]{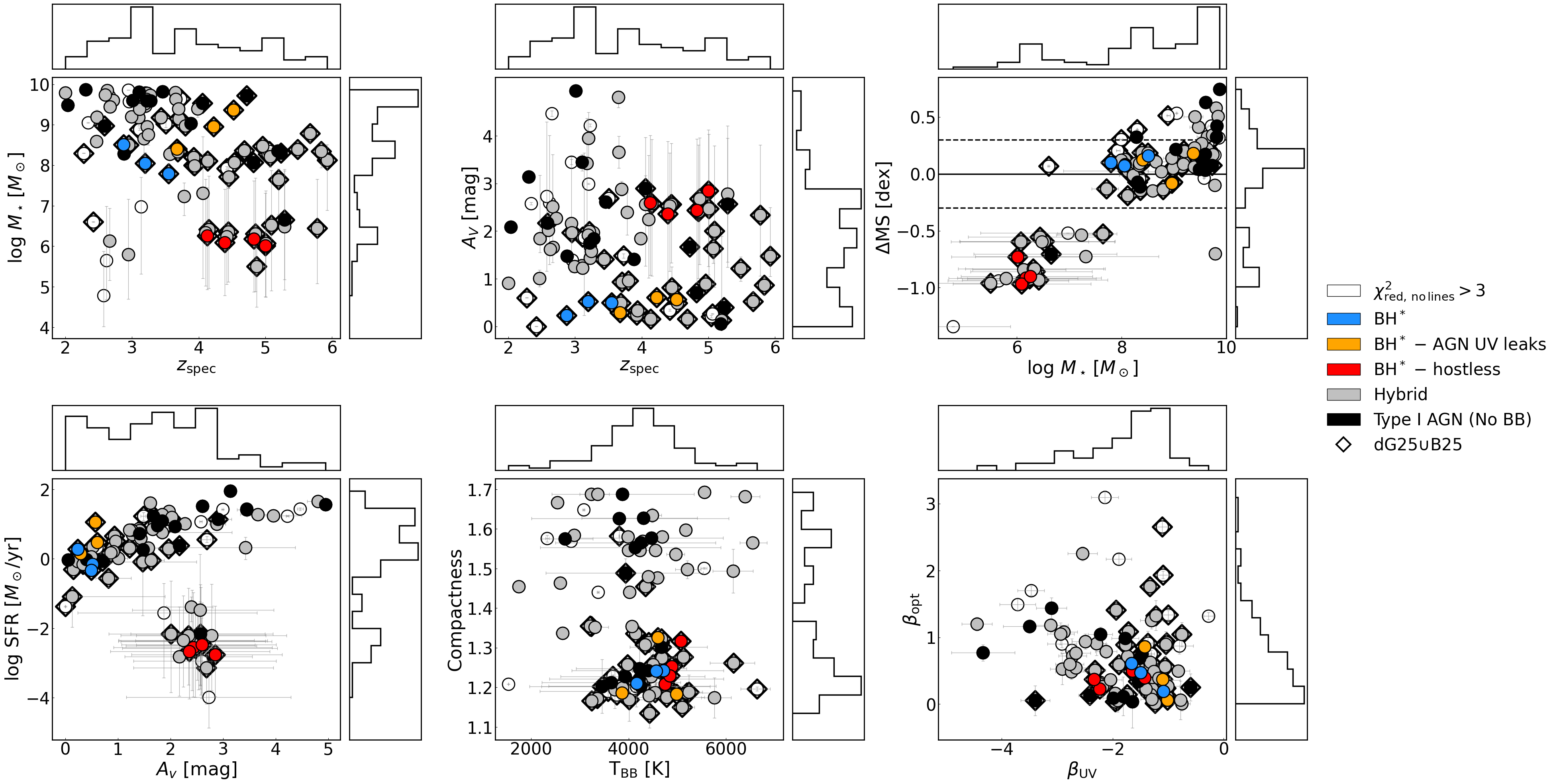}
    \caption{Properties of our sample obtained following \mm\, using broad priors (``AGN-agnostic'' approach). From left to right and top to bottom: stellar mass $M_\star$ vs $z_{\mathrm{spec}}$, dust attenuation $A_V$ vs $z_{\mathrm{spec}}$, distance to the Main Sequence $\Delta$MS vs $M_\star$, star formation rate vs $A_V$, compactness vs BB temperature, and optical vs UV observed slopes. In every plot, we represent $\chi^2_{\mathrm{red}}>3$ fits with empty markers, whereas filled markers correspond to $\chi^2_{\mathrm{red}}<3$ fits (81 LRDs). Blue denotes galaxies in the BH$^*$ class, orange highlights LRDs in the BH$^*$-AGN leaks class, red denotes the BH$^*$-hostless class, Hybrid objects are represented in silver, and Type I AGN systems are plotted in black. Galaxies in common with \dg$\cup$\bb\, are highlighted with diamond markers. We include histograms in every panel, derived from the total sample made up of 99 LRDs. $\Delta$MS values are offsets from the star-forming Main Sequence relation of \citet{Merida2025_ms} fits. The dashed lines denote the typical MS scatter, $\Delta$MS~$\sim0.3$~dex.}
    \label{fig:properties}
\end{figure*}

\begin{itemize}
    \item {\textbf{JADES-12402}}
\end{itemize}

The galaxy showcasing the BH$^*$ class was identified in the JADES fields. It is located at $z=3.1927$ and shows a median S/N of 13.8, as reported by DJA. The parameters defining this LRD resemble those found for \textit{The Cliff} in \mm, classified as BH$^*$ in this work. Our results point to a low-mass, star-forming, metal-poor host subject to a moderate level of dust attenuation ($\sim0.5$~mag), and a temperature of the envelope $\sim4,700$~K. This LRD was also studied in \bb\, and \dg. \bb\, reported values of $\log\,M_\star/M_\odot=8.57\pm0.89$ and $A_V=0.39\pm0.17$~mag, which are compatible with our results within uncertainties. \dg\, found a modified-BB temperature of $4097_{4021}^{4186}$~K, $\sim600$~K colder than our results, although both temperatures are consistent with the theoretical expectations from the BH$^*$ model.

\begin{itemize}
    \item {\textbf{NEXUS-23192}}
\end{itemize}

The LRD showcasing the BH$^*$-AGN leaks class is from NEXUS. It is located at $z=4.5210$ and its median S/N is 11.1. In this case, the host is more massive ($\sim10^{9.4}\,M_\odot$) than \textit{The Cliff}. The level of dust attenuation matches the BH$^*$ example, with a slightly warmer envelope temperature of $\sim5,000$~K. This LRD was included in \dg, but no temperature is reported by these authors, as their analysis of the envelope was only performed on LRDs at $z<4$.

\begin{itemize}
    \item {\textbf{JADES-10868}}
\end{itemize}

The LRD exemplifying the BH$^*$-hostless class is located in the JADES fields ($z=4.3886$, median S/N = 2.9). The host parameters show how \texttt{Bagpipes} is opting for a negligible stellar contribution while pushing for an AGN + BB solution. The code does this by increasing the dust to eliminate the host contribution to the UV continuum and then reducing the $M_\star$ and SFR$_{100}$ (i.e., the SFR measured on a 100~Myr timescale). The lack of stellar constraints is also reflected in the broad 16th-50th-84th percentile range. In terms of the BB temperature, we retrieve a $\sim4,800$~K value, in line with the estimates for the previous categories. This LRD was studied in \dg, who reported a modified-BB temperature of $2,758_{2,533}^{2,990}$~K, much lower than our value and far from the typical range of temperatures expected from the BH$^*$ theory (see Sect.~\ref{sec:intro}). 

\begin{itemize}
    \item {\textbf{NEXUS-3713}}
\end{itemize}

As an example of a \textit{Hybrid} object, which constitutes the most numerous class in our sample, we show an LRD from NEXUS, located at $z = 3.2151$, with a median S/N of 5.4. The $M_\star$ is larger in this source than in the other examples, as well as the $A_V$, which in this case is not an artifact produced by the code to remove the stellar component. Rather, it is a mechanism that suppresses the host's contribution to the UV, which is in turn mainly powered by the AGN. The BB plays a minor role in the optical continuum, with the stellar and AGN components being the primary drivers instead. The BB peak is further away from the ``V'' shape inflection point, resulting in a lower temperature of $\sim4,000$~K. This value is consistent with the temperature reported in \dg, of $4,205_{3,985}^{4,597}$~K.

\begin{itemize}
    \item {\textbf{NIRSpec GTO-Wide-4273}}
\end{itemize}

Lastly, we show an example of a Type I AGN LRD, in which a BB component is not required by the fit. This LRD is from the NIRSpec GTO-Wide ($z=4.0539$, median S/N = 3.2) and corresponds to a massive, star-forming, and dusty galaxy. In this case, the temperature of the BB is not being constrained, as highlighted by the broad 16th-50th-84th percentile range and the low normalization factor. This object was also studied in \bb, who reported values of $\log\,M_\star/M_\odot=8.39\pm0.74$ and $A_V = 0.01\pm0.13$ mag, depicting a dust-free and $\sim1$~dex less massive host. The temperature reported in \dg\, is $5,070_{4,373}^{5,785}$~K.

\subsection{Physical properties}
\label{sec:statistics}

The properties derived following the methodology of \mm\, for the 99 galaxies in our sample are displayed in Fig.~\ref{fig:properties}. Hereafter, we will focus on the LRDs with statistically reliable fits based on the $\chi_{\mathrm{red}}^2$ values (i.e., 81 galaxies out of the total sample of 99 LRDs). A summary of their median properties, distinguishing between classes, can be consulted in Appendix D.

The evolution of the $M_\star$ with $z$ shows that more massive objects appeared at more recent times, with $M_\star>10^{9}\,M_\odot$ retrieved mostly at $z\lesssim4$. The median $M_\star$ of the sample is $\sim10^{8.5}\,M_\odot$, with the majority of more massive galaxies corresponding to Hybrid and Type I AGN-like systems. There is also a subpopulation of galaxies displaying $M_\star\sim10^6\,M_\odot$, which are systems in which the stellar component is not being constrained by the code. 

Bearing in mind that only the Hybrid and Type I AGN classes are sufficiently populated, we see that the objects of the BH$^*$ class, represented in blue, lie at $z<4$ whereas the LRDs in the other quasi-BH$^*$ categories (AGN leaks and hostless, in orange and red) are located at $z\gtrsim4$. The median $M_\star$ of these LRDs, not considering the hostless class, is $\sim10^{8.5}\,M_\odot$, with LRDs in the BH$^*$-AGN leaks class showing $\sim$1~dex larger $M_\star$ than objects in the BH$^*$ category.
Type I AGN-like systems and LRDs in the Hybrid category are spread across all $M_\star$ and $z$, with Type I AGN-like galaxies showing median $M_\star$ values of $\sim10^{9.5}\,M_\odot$ and Hybrid LRDs $M_\star\sim10^{8.5}\,M_\odot$.

In terms of dust attenuation, $A_V>1$~mag values correspond to objects in the Hybrid and Type I AGN categories (in the BH$^*$-hostless class, dust is used as an artifact to remove the stellar contribution to the fit). Objects in the BH$^*$ and BH$^*$-AGN leaks classes display a median $A_V\sim0.5$~mag, in line with the results presented in \mm\, for \emph{The Cliff}.

We also studied the star-forming nature of these galaxies, comparing it with the secular evolution of galaxies on the star-forming Main Sequence (MS). This calculation is based on the MS fits from \citet{Merida2025_ms} for low-mass galaxies. Let us note that these fits were derived using a mass-complete sample that reaches down to $10^8\,M_\odot$ at $2<z<2.5$ ($10^{8.6}\,M_\odot$ at ${5<z<7}$) and thus estimates for lower-mass galaxies are based on extrapolations of these fits.
On average, our sample is consistent with secular evolution (median $\Delta$MS~$\sim0.1$~dex, with $\Delta$MS being the distance to the MS). LRDs in the BH$^*$-hostless category fall below the MS ($\Delta$MS~$\sim-1$~dex, although neither SFR nor $M_\star$ are being well constrained by the code), and some Hybrid and Type I AGN objects lie below and above the typical MS scatter (i.e., $\Delta$MS~$\sim0.3$~dex). Galaxies with the highest SFR values also correspond to objects with the largest $A_V$ values, pointing to obscured SFR in those systems.

In terms of the envelope temperature, our sample is concentrated around $\sim4,400$~K, with BH$^*$-hostless LRDs showing slightly warmer temperatures ($\sim4,900$~K). When analyzing compactness, only sources in the Type I AGN and Hybrid classes are found close to the compactness limit (i.e., ${f(0.5\arcsec)/f(0.3\arcsec)= 1.7}$, with $f$ being the flux at $F444W$). Conversely, most of the sources exhibit ${f(0.5\arcsec)/f(0.3\arcsec) < 1.4}$ values.

Our classification does not yield clear delimitations or trajectories in the $\beta_{\mathrm{opt}}\,\mathrm{vs}\,\beta_{\mathrm{UV}}$ diagram. Only some Type I AGNs and Hybrid objects are located farther from the bulk of the points, which are concentrated at $\beta_{\mathrm{opt}}<1$.
 
These results demonstrate that, based on the analysis of the continuum, the physical properties of LRDs can vary significantly, with only a few LRD hosts matching the characteristics of a low-mass, metal-poor, low-dust galaxy. Furthermore, only a few LRDs satisfy the overall requirements of the BH$^*$ model. In Sect.~\ref{sec:discussion_1} we explore the physical implications of our solutions, while in Sect.~\ref{sec:discussion_2} we modify our method to force a BH$^*$ scenario across the entire the sample.

\section{Discussion}
\label{sec:discussion}

\subsection{Viability of the current solution}
\label{sec:discussion_1}

As mentioned in Sect.~\ref{sec:method}, our method involves using broad priors and providing all the necessary ingredients to produce the BH$^*$ scenario, should it be supported by the data,  while still allowing \texttt{Bagpipes} to converge on a different, non-BH$^*$ best-fitting solution. This means that we are not imposing any particular physical scenario, and the solutions are data-driven only.

In a small number of cases, \texttt{Bagpipes} indeed naturally yields a BH$^*$ configuration ($\sim5,000$~K BB emission and a low-mass, low-$A_V$ host galaxy). However, most of the LRDs in our sample do not fully meet the requirements of the BH$^*$ model, yet we still obtain statistically robust fits. 
Nevertheless, mathematical robustness does not imply physical robustness; it is important to compare our results with additional observables. For example, past attempts to fit the UV-to-optical emission of LRDs based on composite models consisting of stars and AGN emission, with no gas envelope, yielded good fits in some LRDs (e.g., \citealt{Merida2025}, \citealt{Tripodi2024}). However, the high $A_V$ values obtained in some of these studies (and high $M_\star$ in some cases) often exceeded those expected based on the observed dust masses in LRDs (see \citealt{Casey2024}, \citealt{Setton2025}).
In our particular case, some of the solutions presented in Sect.~\ref{sec:results} also require high $A_V$ values, which would likewise be difficult to reconcile with these dust mass observations.  

Furthermore, our model yields different AGN contributions to the UV, ranging from 0 to 100\% of the total UV flux. However, if the host galaxy were resolved in the UV (see \citealt{Chen2025}, \citealt{Rinaldi2024}, \citealt{Cloonan2026}), this emission could be attributed to the host galaxy. The presence of an extended emission in the UV would therefore call into question a potential AGN origin of the UV emission.

Finally, the classes defined in Sect.~\ref{sec:demographics} are based on the analysis of the continuum alone, not including the information encoded in the emission lines. Consequently, these continuum decompositions of LRDs may not be related to the emission-line properties, meaning they may not represent distinct physical types of LRD.

In this subsection, we examine the validity of our solutions in light of all these physical considerations.

\subsubsection{Dust attenuation}
\label{sec:discussion_1_2}

From the BH$^*$ perspective, LRD hosts are dust-free/low-$A_V$ systems. In \mm, we inferred dust attenuation values for \textit{The Cliff} from its UV-to-optical SED, obtaining $A_V \sim 0.5-1$~mag depending on the adopted attenuation law (see also \pg). 
Based on the infrared (IR)-to-submillimeter emission of LRDs, \citet{Chen2025b} estimated $A_V$ upper limits for these objects using data from the JWST Mid-Infrared Instrument, \textit{Herschel}, and the Atacama Large Millimeter/submillimeter Array. They obtained values of ${A_V \leq 1-1.5}$~mag (depending on the dust attenuation law selected) for the brightest LRDs to date, A2744-45924 \citep{Labbe2024environment} and RUBIES-BLAGN-1 \citep{Wang2025}, as well as for stacked SEDs from a large sample of LRDs. Consequently, some level of dust attenuation is permitted from both the UV-to-optical and IR-to-submillimeter perspectives, but this quantity is subject to stringent constraints.  

Looking at Fig.~\ref{fig:properties}, we see that the objects from the BH$^*$ and BH$^*$-AGN leaks categories, as well as some Type I AGN and Hybrid LRDs, agree well with the upper limits derived by \citet{Chen2025b}. In the BH$^*$-hostless case, the $A_V$ is a mathematical artifact of the code, but there are many Hybrid and Type I AGNs LRDs that exhibit $A_V>1.5$~mag, pointing to potentially nonphysical solutions.

\begin{figure*}[!htp]
    \centering
    
    \begin{subfigure}{\textwidth}
    \centering
    \includegraphics[width=.33\linewidth]{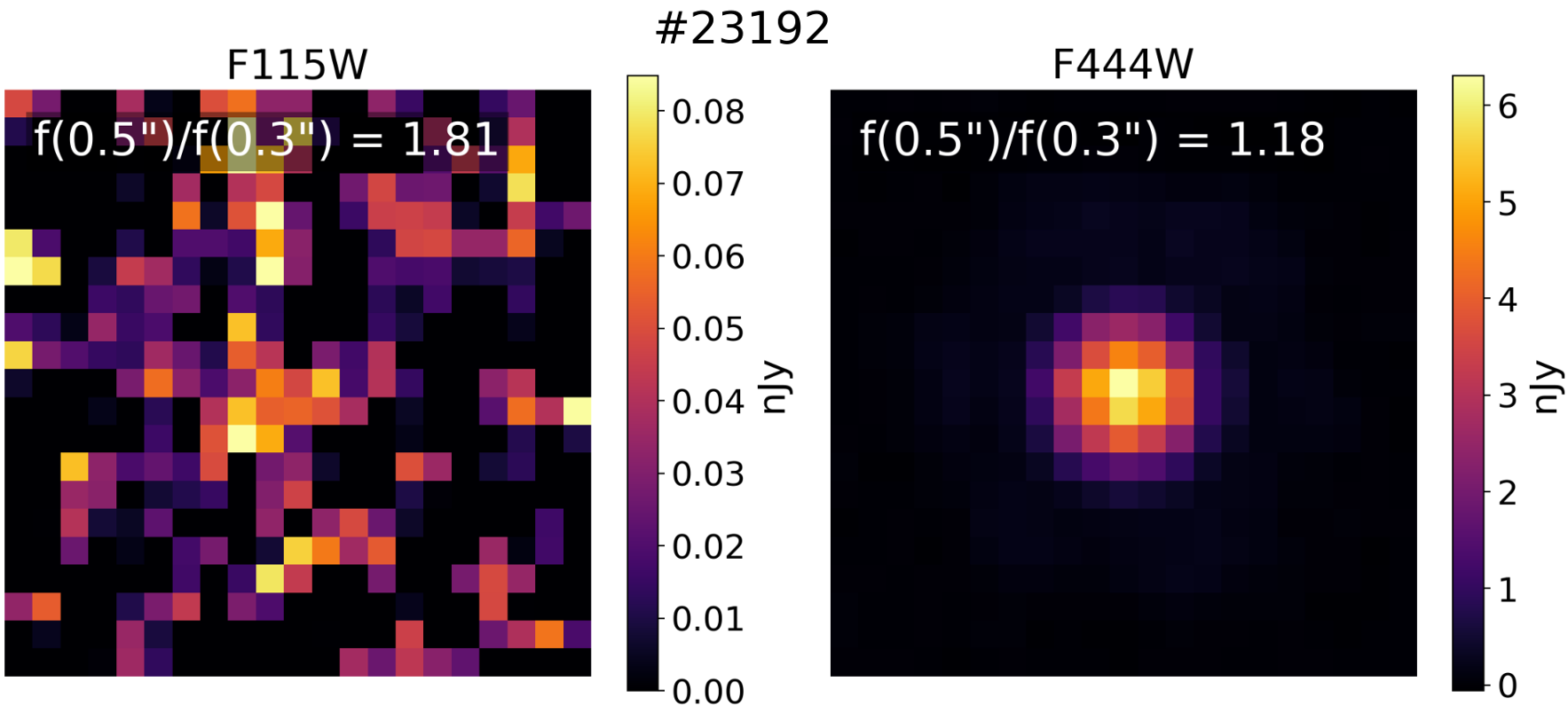}
    \includegraphics[width=.33\linewidth]{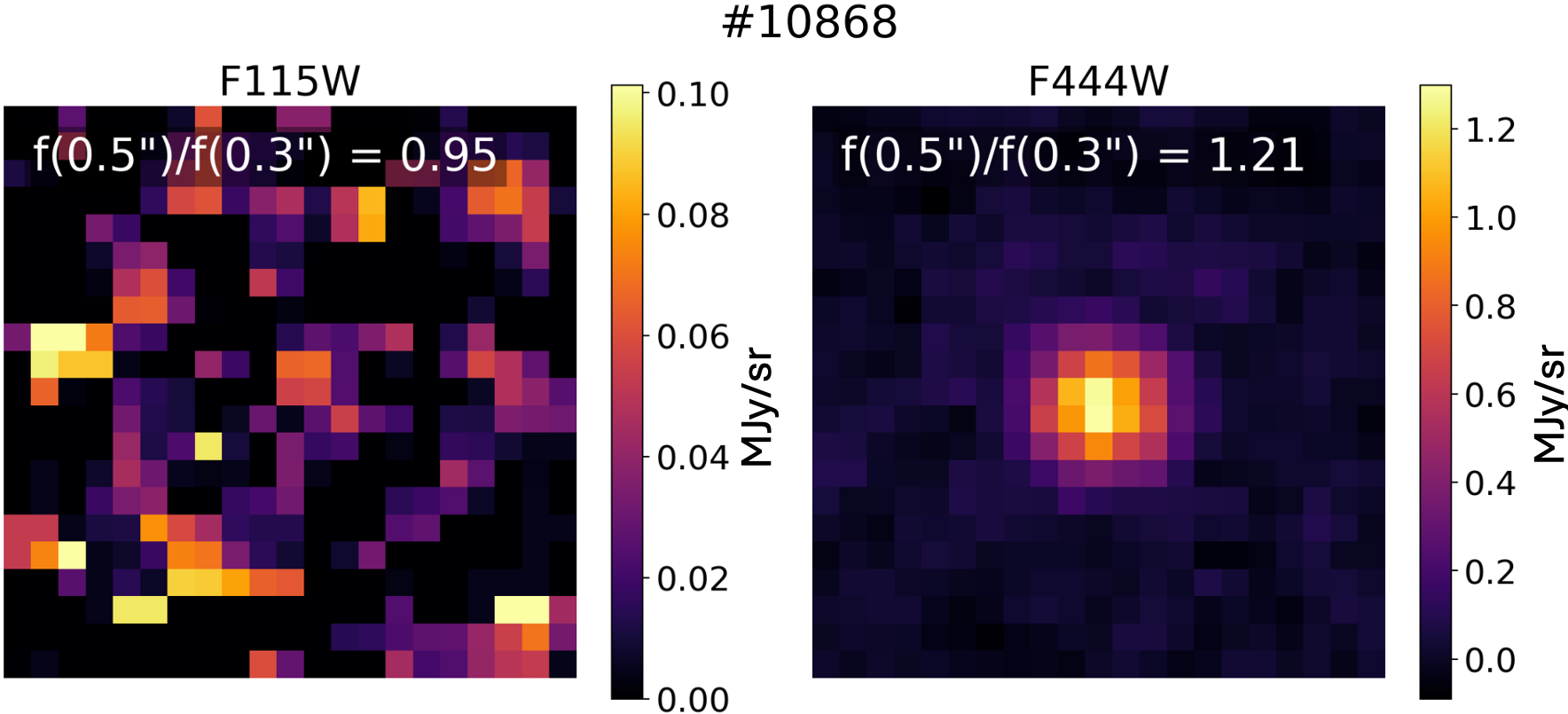}
    \includegraphics[width=.33\linewidth]{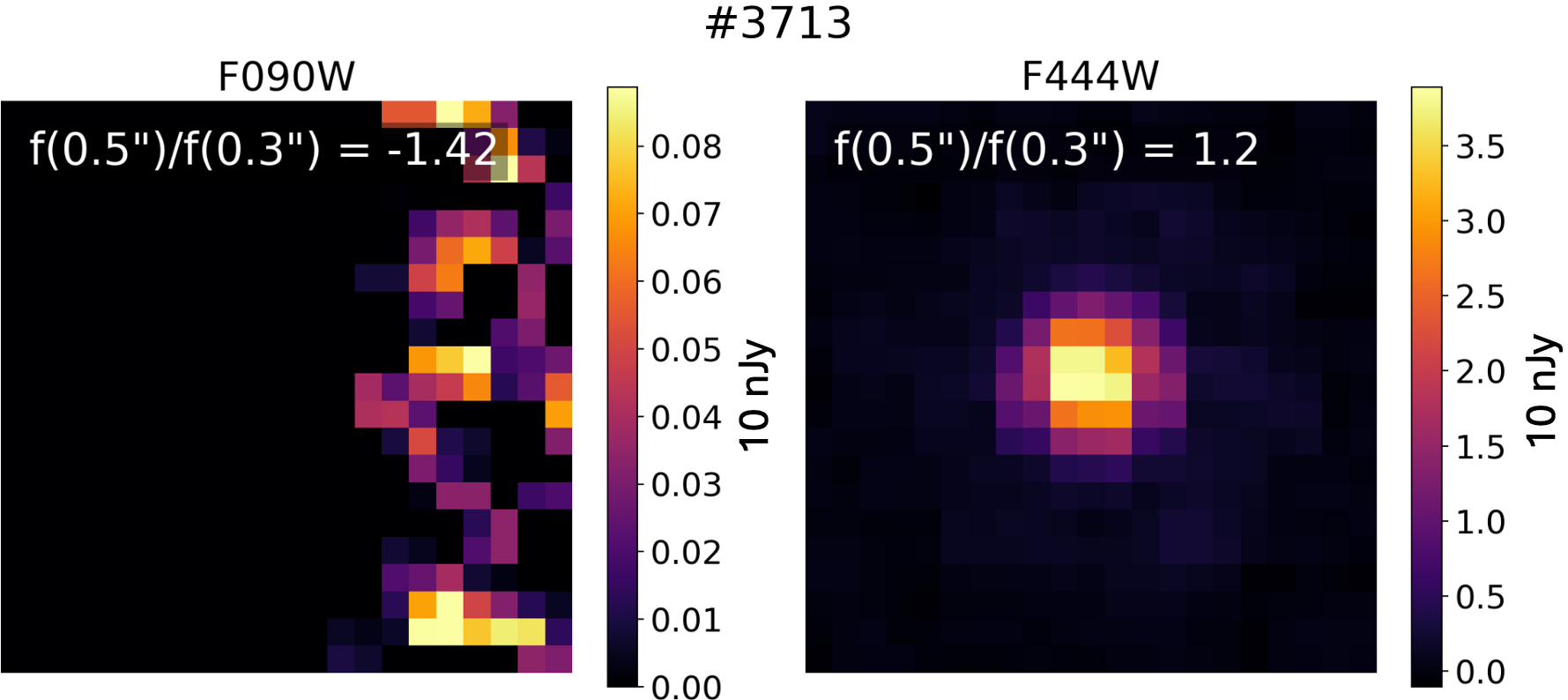}
    \end{subfigure}

    \begin{subfigure}{\textwidth}
    \centering
    \includegraphics[width=.7\linewidth]{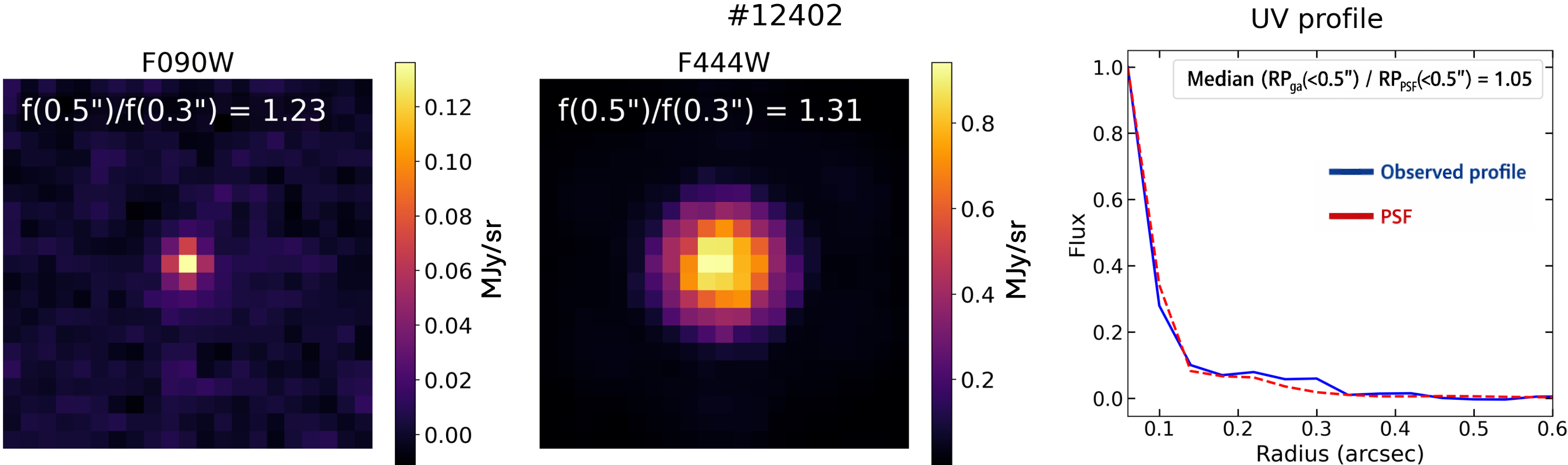}
    \end{subfigure}
    
    \begin{subfigure}{\textwidth}
    \centering
    \includegraphics[width=.71\linewidth]{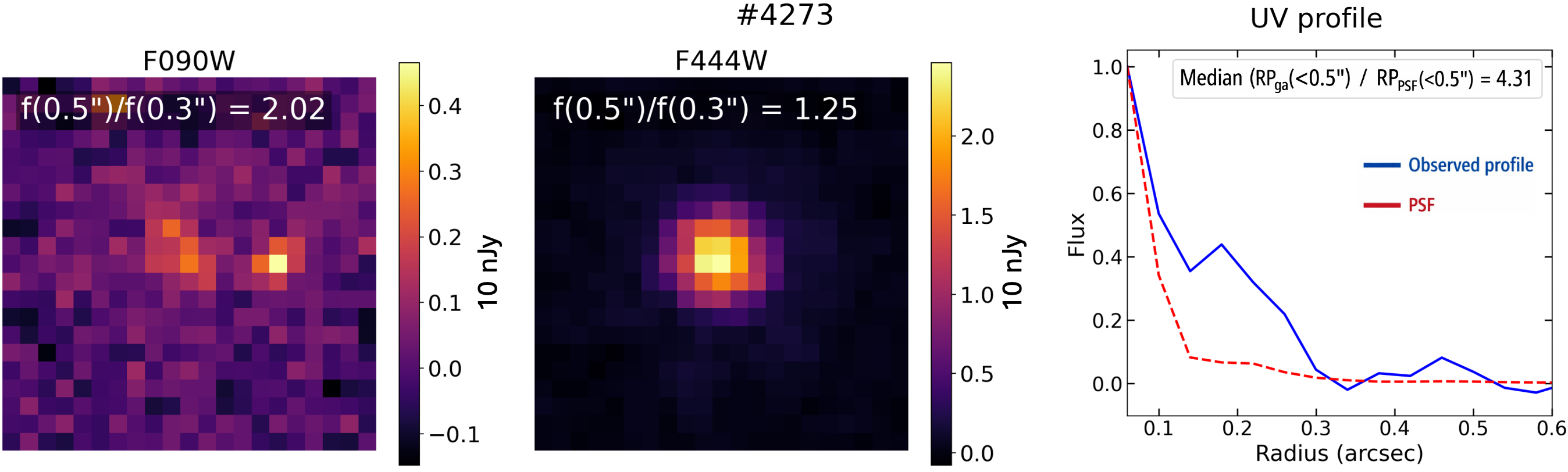}
    \end{subfigure}

    \caption{$0.8\times0.8$ arcsec$^2$ cutouts in the rest-frame UV ($F090W$ or $F115W$) and $F444W$ of the LRDs displayed in Fig.~\ref{fig:example_classes}. In the legend, we show the ratio of the flux measured in that band using 0.5 and 0.3\arcsec\, apertures. The first row shows the results for LRDs \#23192, \#10868, and \#3713, which belong to the BH$^*$-AGN leaks, BH$^*$-hostless, and Hybrid categories, respectively. The second and third row, showing LRDs \#12402 (BH$^*$) and \#4273 (Type I AGN), also include the UV radial profile (blue) compared to the PSF (red), as well as the median ratio of these two curves, as measured for radii $<0.5$\arcsec. While the UV morphology is sometimes difficult to measure due to noise, we can identify structure in some objects. In a minority of cases, we see hints of a potential resolved galaxy host (e.g., \#4273).}
    \label{fig:UV_profiles}
\end{figure*}

However, note that \citet{Chen2025b} computed their $A_V$ upper limits by considering different dust distribution parameters: $n_0$ and $\gamma$. Here, $n_0$ is the density at the inner edge of the radial density profile, and $\gamma$ is the power-law index. In other words, they used a range of values to control the normalization and concentration of the dust density profile. Their ${A_V \leq 1-1.5}$~mag upper limit is based on the assumption of concentrated density profiles ($\gamma\geq 1$), as the high-density regime allows for significantly lower dust masses. However, less restrictive $A_V$ upper limits are obtained for flat density profiles ($\gamma \sim 0$) with $n_0 \sim 10-30\, \mathrm{cm}^{-3}$. 

While these alternative values for the $A_V$ upper limits are still incompatible with the dust-obscured AGN scenario \citep{Wang2024} (i.e., a scenario without a gas envelope, which would require $A_V\sim3$ mag), they are consistent with $A_V\sim1.4$~mag using the SMC law, ${A_V\sim1.9}$~mag using a Milky Way dust law, and $A_V\sim2.4$~mag using an Orion Nebula dust law in the case of RUBIES-BLAGN-1. Therefore, depending on the dust law and dust configuration, $A_V$ values such as those reported in Fig.~\ref{fig:properties} could be physically realistic in some cases.

Nevertheless, \citet{Chen2025b} caution that such profiles may represent unphysical configurations. They point out that, if the density profile connects the BLR and dust sublimation scales, flat profiles are unlikely. Therefore, alternative scenarios invoking lower $A_V$ values should be explored for the highly attenuated LRDs in our sample.

\subsubsection{UV morphology}
\label{sec:discussion_1_1}

We investigated the UV morphology of the potential LRDs' galaxy hosts to explore the probable origin of the UV emission.
If the host galaxy were resolved, the UV emission would predominantly originate from the stellar component, ruling out an AGN origin (or at least the AGN as the primary mechanism). However, an unresolved host would not provide conclusive evidence, as in that case the galaxy may simply be too faint to be detected.

For this exercise, for each LRD in the sample, we selected the NIRCam band closest to the rest-frame UV range and compared the normalized radial profile of the LRD within a radius of 0.5\arcsec\, with that of the point spread function (PSF). The PSFs were taken from the library of simulated NIRCam PSFs provided by the Space Telescope Institute and produced using the Space Telescope PSF package \citep{Perrin2014}. These are not PSFs extracted empirically for each image. However, they should suffice here as our aim is not to carry out a detailed study of UV morphologies, but rather to detect potential emission that is incompatible with a PSF-like profile.

Figure~\ref{fig:UV_profiles} shows cutouts in the rest-frame UV of the objects described in Sect.~\ref{sec:demographics} and displayed in Fig.~\ref{fig:example_classes}. Overall, we did not find UV coverage for some LRDs, such as object \#3713, classified as Hybrid. In other cases, such as in objects \#23192 and \#10868 belonging to the BH$^*$-AGN leaks and -hostless classes, respectively, the image is too noisy. In neither of these cases can we rule out any UV origin. 

We could study the UV radial profiles of objects \#12402 and \#4273, belonging to the BH$^*$ and Type I AGN classes, respectively. The profile of object \#12402 is consistent with the PSF and the unresolved nature of the UV emission, with $f(0.5\arcsec)/f(0.3\arcsec)=1.23$ and $f$ being the flux in the rest-frame UV. Therefore, stellar and stellar+AGN origins remain degenerate solutions for this LRD. 

LRD \#4273 exhibits signs of complex and extended UV morphology ($f(0.5\arcsec)/f(0.3\arcsec)=2.02$ in $F090W$), indicating the presence of two potential clumps or galaxies. The rightmost component corresponds to either a companion/interacting galaxy (see \citealt{Chen2025}, \citealt{Merida2025}, \citealt{Baggen2026}, \citealt{Golubchik2025}) or a projected foreground source. It is quite challenging to derive a robust photo-$z$ for this potential companion, as the small projected separation ($\sim0.2$\arcsec, corresponding to $\sim1.5$~kpc at $z=4.05$) likely leads to contamination of the longer-wavelength photometry. Nevertheless, at $z = 4.05$ galaxies are  $V$-band dropouts and, in this case, both clumps remain undetected in the $F435W$ band of the \textit{Hubble Space Telescope}. Consequently, the observed UV emission may arise from the combined contribution of the two clumps or galaxies associated with the LRD system.

Under this interpretation, a dominant stellar contribution to the UV emission appears more plausible. However, the best-fitting solution for this object, shown in Fig.~\ref{fig:example_classes}, attributes most of the UV flux to the AGN continuum, suggesting that this solution may not provide a physically realistic description of this LRD and that alternative approaches should be considered. 

Note, however, that most of the objects in our sample remain undetected or unresolved in the rest-frame UV. We found three additional LRDs with potential host signatures, \#47509, \#23438 and \#53757, also included in \dg$\cup$\bb\, and classified as Hybrid in Sect.~\ref{sec:demographics}.
To delve deeper into alternative solutions likely invoking lower $A_V$ values and reducing the role played by the AGN continuum in the fit, we will explore a modification of our current fitting method in Sect.~\ref{sec:discussion_2}.

\subsection{Emission-line diagnostics}
\label{sec:elines}

The presence of absorption in the Balmer lines, exponential wings, and extreme Balmer breaks in a significant fraction of LRDs was key in supporting the BH$^*$ scenario. Our analysis does not currently consider the information encoded in the emission lines; thus, the connection between the classes defined in Sect.~\ref{sec:demographics} and the emission line properties of LRDs is yet to be determined. A full analysis of the emission-line properties is out of the scope of this work. However, in this section, we briefly discuss these categories in relation to the [OIII] line emission and the Balmer break strength. The origin of the [OIII] emission is usually attributed to star-forming regions in the host galaxy. The Balmer break strength correlates positively with the optical continuum luminosity (i.e., the BB-like emission) \citep{deGraaf_review}. Comparison of both properties should hence allow us to investigate the relation between the host and gas envelope contributions to the LRD spectra.

The [OIII] luminosity was derived from the integrated flux of the [OIII]$\lambda 4959,\lambda5007$ doublet. We fitted the emission lines with Gaussian profiles using the \texttt{lmfit} \texttt{Python} package \citep{Newville2014} and adopting the theoretical flux ratio $F_{5007}/F_{4959} = 2.98$. The contribution of the broad H$\beta$ component to the [OIII] doublet was taken into account by modeling a broad component simultaneously with the [OIII] lines, thus removing potential contamination due to H$\beta$. Uncertainties were estimated from Monte Carlo realizations of the best-fitting profiles.

The Balmer break strength was measured as the ratio of the mean flux densities in two emission-line-free rest-frame wavelength windows, $D_B=F_{\nu,4225}/F_{\nu,3565}$, corresponding to $4160-4290\,\AA$ and $3500-3630\,\AA$, respectively (see \citealt{Antwi-Danso2025}). The errors were estimated through Monte Carlo propagation of the flux uncertainties.

\begin{figure}[htp]
    \centering
    \includegraphics[width=0.9\linewidth]{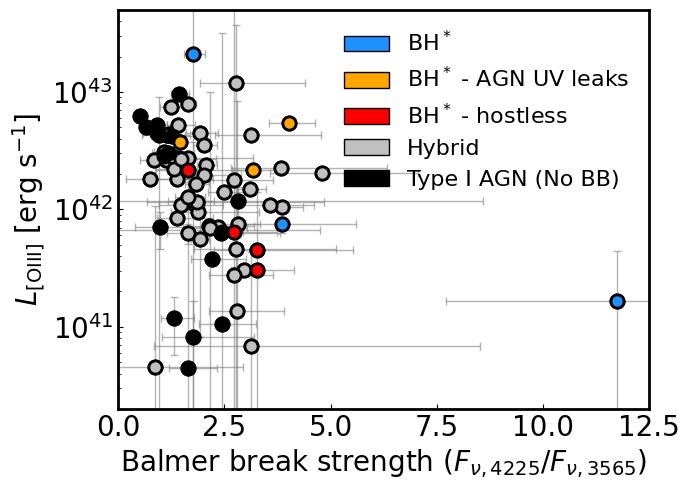}
    \caption{[OIII] luminosity vs. the Balmer break strength $D_B$. Markers are color-coded according to the classes defined in Sect.~\ref{sec:demographics}. The [OIII] luminosity shows an anticorrelation with $D_B$, as also reported in previous works such as \citet{deGraaf_review}.}
    \label{fig:spectral}
\end{figure}

\begin{figure*}[htp]
    \centering
    \includegraphics[width=\linewidth]{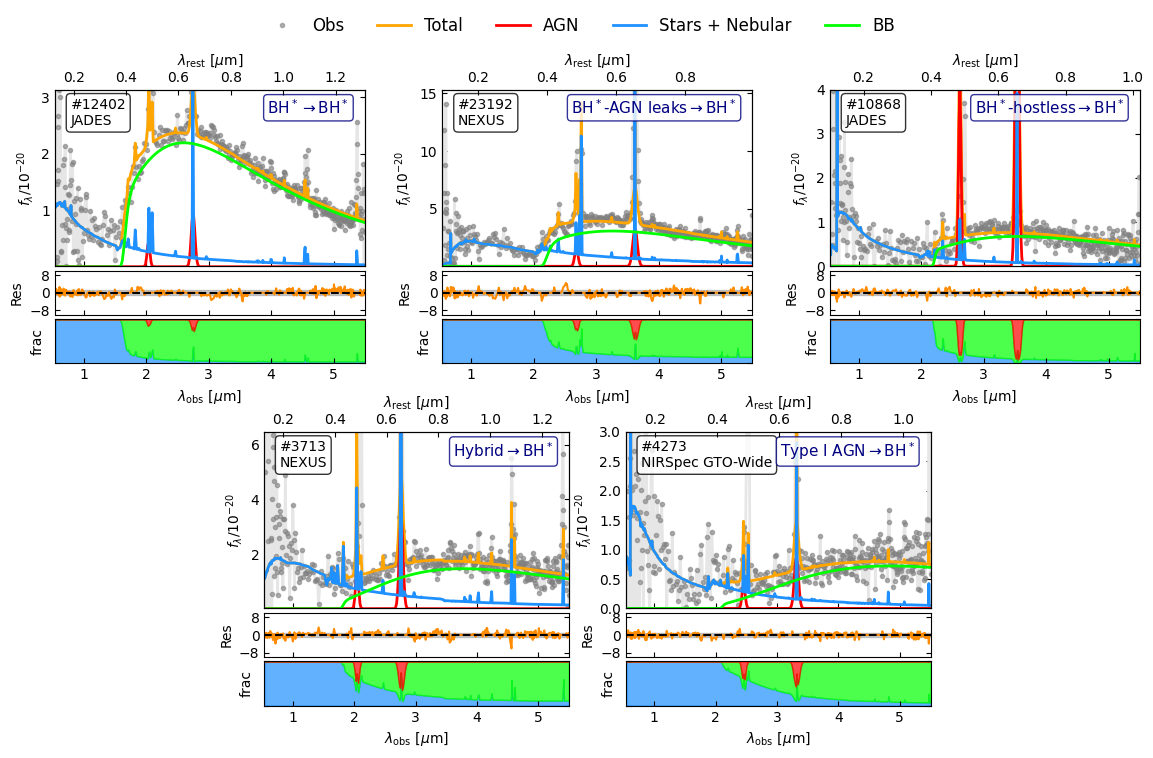}
    \caption{Best-fitting models from \texttt{Bagpipes} based on the continuum of the LRD subset displayed in Fig.~\ref{fig:example_classes} obtained after suppressing the AGN continuum in the fit. See Fig.~\ref{fig:example_classes} for a full description of the markers and color codes shown here. In the legend, we specify the classes (see Table~\ref{tab:classes}) based on the AGN-agnostic method. The new class that results from suppressing the AGN in the fit is indicated with an arrow. In this case, all the objects belong to the BH$^*$ class when suppressing the AGN in the code.}
    \label{fig:example_classes_forced}
\end{figure*}

\begin{figure*}[htp]
    \centering
    \includegraphics[width=\linewidth]{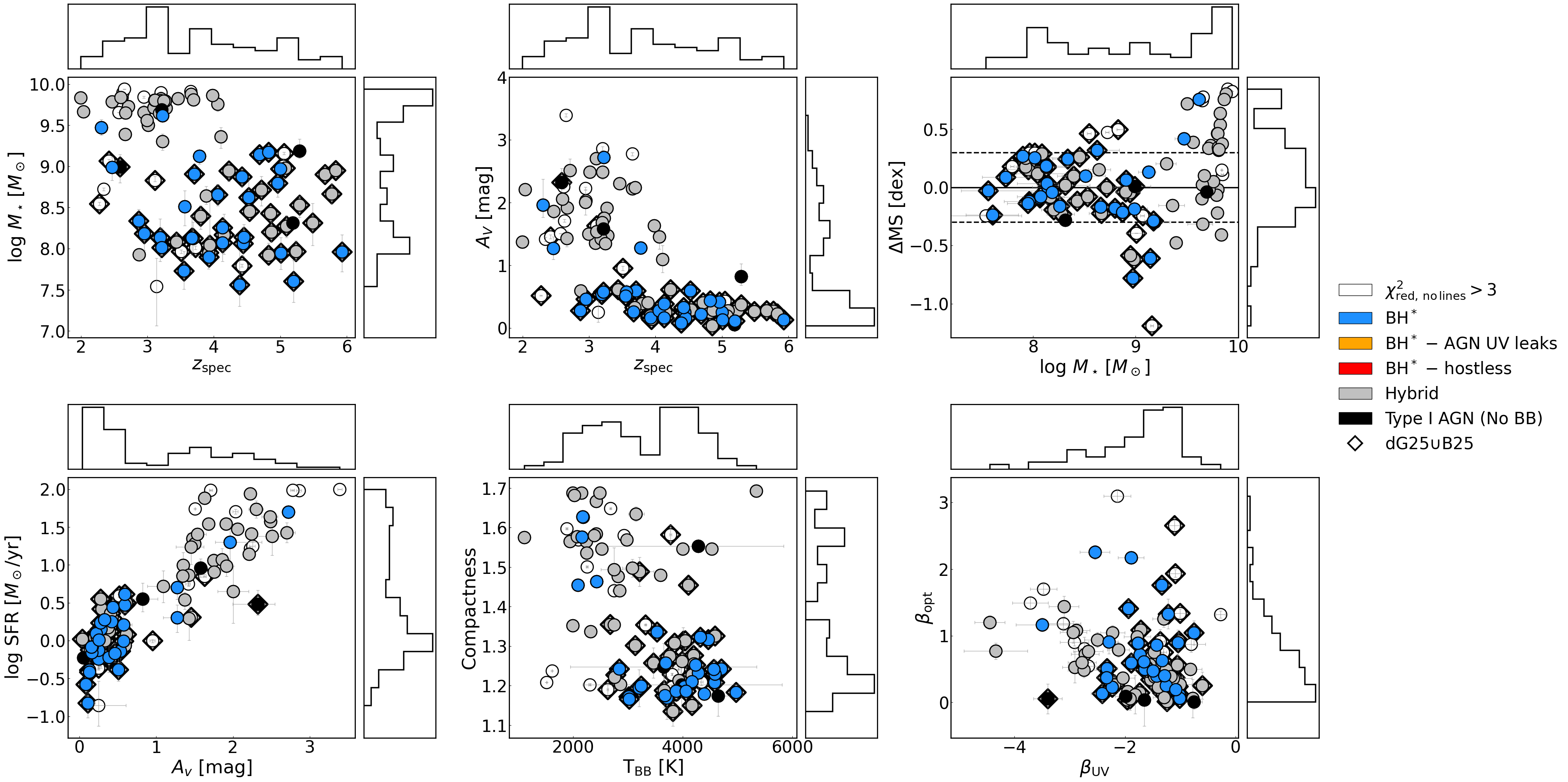}
    \caption{Properties of our sample after suppressing the AGN continuum. From left to right and top to bottom: stellar mass $M_\star$ vs $z_{\mathrm{spec}}$, dust attenuation $A_V$ vs $z_{\mathrm{spec}}$, distance to the Main Sequence $\Delta$MS vs $M_\star$, star formation rate vs $A_V$, compactness vs BB temperature, and optical vs UV observed slopes. See Fig.~\ref{fig:properties} for a full description of the markers and color codes shown here.}
    \label{fig:properties_forced}
\end{figure*}

Figure~\ref{fig:spectral} displays the [OIII] luminosity vs. $D_B$ distinguishing between the classes defined in Sect.~\ref{sec:demographics}. As also reported in \citet{deGraaf_review}, the [OIII] luminosity shows an anticorrelation with $D_B$ (albeit with some outliers from the Type I AGN and Hybrid classes). The outlier at $D_B>10$ is \textit{The Cliff}, whose Balmer strength is uncertain but much higher than the average values of $D_B$ seen in the rest of the LRDs in the sample and consistent with \citet{deGraaf2025} within uncertainties.

Type I AGNs show lower values of $D_B$ and are spread across a wide range of [OIII] luminosities. Hybrid LRDs show, on average, higher $D_B$ values compared to objects in the Type I AGN class. The BH$^*$ and quasi-BH$^*$ categories comprise a small number of LRDs, which makes it difficult to investigate trends in this diagram. However, LRDs in the BH$^*$-AGN leaks class show high [OIII] luminosities and $D_B$ ranging $\sim2-5$, which could point to a contribution of the AGN to the [OIII] luminosity, added to the stellar emission (see \citealt{Sok2026}). Conversely, LRDs in the BH$^*$-hostless class show lower [OIII] luminosities and are spread across a similar $D_B$ range. A lower [OIII] luminosity could be due to the absence of photoionization from the galaxy host.

Lastly, there is not a common behavior among the LRDs in the BH$^*$ class, spanning a wide range of [OIII] luminosities and $D_B$ values.
A larger sample of LRDs in the BH$^*$ and quasi-BH$^*$ categories is required to determine whether this continuum-based classification corresponds to distinct physical populations. If confirmed, the emerging differences in the [OIII] emission would suggest that the continuum properties can trace variations in the physical conditions of the ionized gas.

\subsection{Forcing a BH$^*$ scenario}
\label{sec:discussion_2}

The original method presented in \mm\, is based on the assumption of broad priors for all the components. This enables \texttt{Bagpipes} to identify the best-fitting model based on the data, and allows potential AGN continuum contribution to the model flux. In Sect.~\ref{sec:results}, we show that following this permissive approach the BH$^*$ is only recovered in a subset of objects. In most cases the solutions do not align with the BH$^*$ framework requirements, yielding instead high levels of dust attenuation and AGN UV leakage in some LRDs. This high level of dust may conflict with the $A_V$ upper limits inferred from the typical IR-to-submillimeter emission of LRDs \citep{Chen2025b}, as well as the potential resolved UV emission detected in some of them (see Sect.~\ref{sec:discussion_1}).

In order to prevent potential AGN leakage and limit $A_V$, in this subsection we repeat the analysis performed in Sect.~\ref{sec:results} with one key difference: we manually suppress the AGN continuum. We do this by lowering the \texttt{f5100A} parameter, imposing a range of [0, 5$\times10^{-24}$] erg/s/cm$^2$/$\AA$ instead. The rest of the priors were left unchanged (see Table~\ref{tab:priors}). 

As a result of this new approach (which we refer to as ``AGN-suppressed''), we obtained 81 fits with $\chi^2_{\mathrm{red}}<3$. We include a comparison of the properties reported in this subsection with those from \dg$\cup$\bb\, in Appendix A. 
Following the criteria listed in Table~\ref{tab:classes}, 28 LRDs are now classified as BH$^*$, 49 belong to the Hybrid class, and 4 correspond to the Type I AGN class. The other two classes, BH$^*$-AGN leaks and -hostless, remain empty by construction of the model. In this case, the objects in the Type I AGN class correspond to galaxies whose continuum is purely stellar, whereas Hybrid objects display an optical continuum that is not dominated by the BB component, but have significant contributions from the galaxy host. 

Figure~\ref{fig:example_classes_forced} shows the LRDs displayed previously in Fig.~\ref{fig:example_classes}, together with the new best-fitting models. Table B.1 shows a full list of their properties (``AGN-suppressed'' column). Best-fitting models for the rest of the objects based on this approach, distinguishing between classes, can be consulted in Appendix E.

The $\chi^2_{\mathrm{red}}$ values obtained for these and the rest of the LRDs in the sample do not differ significantly from those reported in Sect.~\ref{sec:demographics}, making the ``AGN-agnostic'' and ``AGN-suppressed'' solutions statistically degenerate. Additional observational constraints are therefore required to resolve this conflict.

The solution for LRD \#12402, which was also previously classified as BH$^*$, remains consistent with the results presented in Sect.~\ref{sec:demographics}. LRD \#23192, classified as BH$^*$-AGN leaks under the ``AGN-agnostic'' configuration, now belongs to the BH$^*$ class, with a galaxy host that is  $\sim0.8$~dex less massive than before, and that exhibits the same amount of dust attenuation and a similar BB temperature.

Object \#10868, classified as BH$^*$-hostless in Sect.~\ref{sec:demographics}, is fitted with a dust-free $\sim10^{7.6}\,M_\odot$ galaxy host. The temperature in this case is lower by $\sim200$~K. LRD \#3713, which showcased the Hybrid category, now corresponds to a galaxy that's $\sim1.6$~dex less massive than before, with a dust attenuation $A_V\sim0.58$~mag, now in line with a \textit{Cliff}-like level of dust. However, the temperature is $\sim600$~K colder than the value reported in Sect.~\ref{sec:demographics}. Finally, object \#4273, which exemplified the Type I AGN category in Sect.~\ref{sec:demographics} and shows hints of a complex UV morphology (see Sect.~\ref{sec:discussion_1_1}), is now a $\sim1$~dex less massive galaxy subject to low dust attenuation ($A_V\sim0.3$~mag).

In these last three examples (\#10868 BH$^*$-hostless, \#3713 Hybrid, and \#4273 Type I AGN), suppressing the AGN continuum results in lower dust attenuation levels, in line with the numbers discussed in Sect.~\ref{sec:discussion_1_2}. While the BB temperatures in LRDs \#12402 and \#23192 (BH$^*$ and BH$^*$-AGN leaks) remain unchanged when the AGN is suppressed, the temperatures for the objects previously classified as BH$^*$-hostless and Hybrid are lower than those reported in Sect.~\ref{sec:demographics}.

For objects \#3713 and \#4273 (Hybrid and Type I AGN), the stellar component now contributes significantly to the optical continuum close to the Balmer limit. Consequently, under the ``AGN-suppressed'' approach the BB peak shifts to longer wavelengths (i.e., colder temperatures) in these LRDs.

\subsubsection{Physical properties. AGN-agnostic vs AGN-suppressed}

By suppressing the AGN component during the fitting process, the BH$^*$ scenario is effectively enforced. As a consequence, lower $M_\star$ and $A_V$ values, together with colder temperatures, tend to be favored. In this subsection, we examine the properties of the full sample to assess whether these trends become statistically significant when increasing the sample size.

Figure~\ref{fig:properties_forced} presents the inferred properties of the full LRD sample obtained with this ``AGN-suppressed'' approach, analogous to Fig.~\ref{fig:properties}, which is based on the ``AGN-agnostic'' method. A summary of the median properties of the sample using the AGN-suppressed method is also included in Appendix D. 

Examining the diagrams reveals two distinct subpopulations. On the one hand, there are LRDs with $M_\star<10^9\,M_\odot$ that are subject to moderate to low dust attenuations ($A_V\sim1$~mag) and lie within the MS scatter. These are compact sources with BB temperatures around $\sim4,500$~K, showing some scatter towards colder temperatures. This first subpopulation aligns with the expectations of the BH$^*$ model.

On the other hand, there is a subpopulation of massive (${\sim10^{9-10}\,M_\odot}$) and dusty systems ($A_V>1.5$~mag), some of them starbursts. These are also less compact objects. This subpopulation mainly comprises LRDs in the Hybrid category according to the ``AGN-suppressed'' approach, where the BB component is subdominant to the stars in the optical. 
Within this massive and dusty subpopulation, there are also two LRDs now classified as BH$^*$ systems, \#1333 and \#150522, neither of them studied in \dg$\cup$\bb. These LRDs were previously classified as Hybrid and Type I AGN in Sect.~\ref{sec:demographics}. They have complex continua, showing evidence for Paschen Jumps, which makes it difficult to constrain the full optical continuum with a single temperature. 

Paschen jumps in LRDs were also explored in \citet{Sneppen2026}, who report evidence for this feature in a subset of LRDs, consistent with free-bound recombination to hydrogen $n = 3$. In LRDs with the strongest jumps, they found that simple BB emission can reproduce the Brackett continuum or the very bluest Paschen continuum emission. However, it cannot smoothly fit both wavelength regimes, pointing to the need for more sophisticated models.

\begin{figure}[htp]
    \centering
    \includegraphics[width=0.8\linewidth]{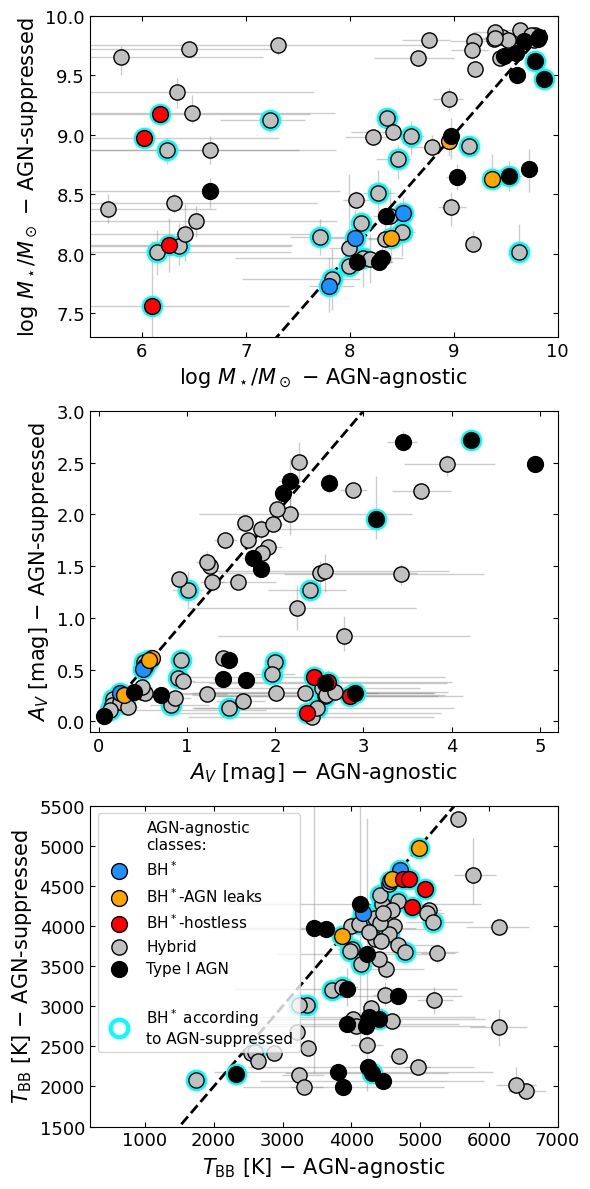}
    \caption{A comparison of the $M_\star$ (top), $A_V$ (middle), and BB temperature (bottom) is shown for the AGN-agnostic and AGN-suppressed methods. Only results in which $\chi^2_{\mathrm{red}}<3$ according to both methods (81 LRDs in total) are included. Markers are color-coded following the classification based on the AGN-agnostic results. Objects that are classified as BH$^*$ systems according to the AGN-suppressed method are encircled in cyan.}
    \label{fig:free_vs_suppressed}
\end{figure}

\begin{figure*}[!htp]
    \centering
    \begin{subfigure}{.56\textwidth}
    \centering\includegraphics[width=\linewidth]{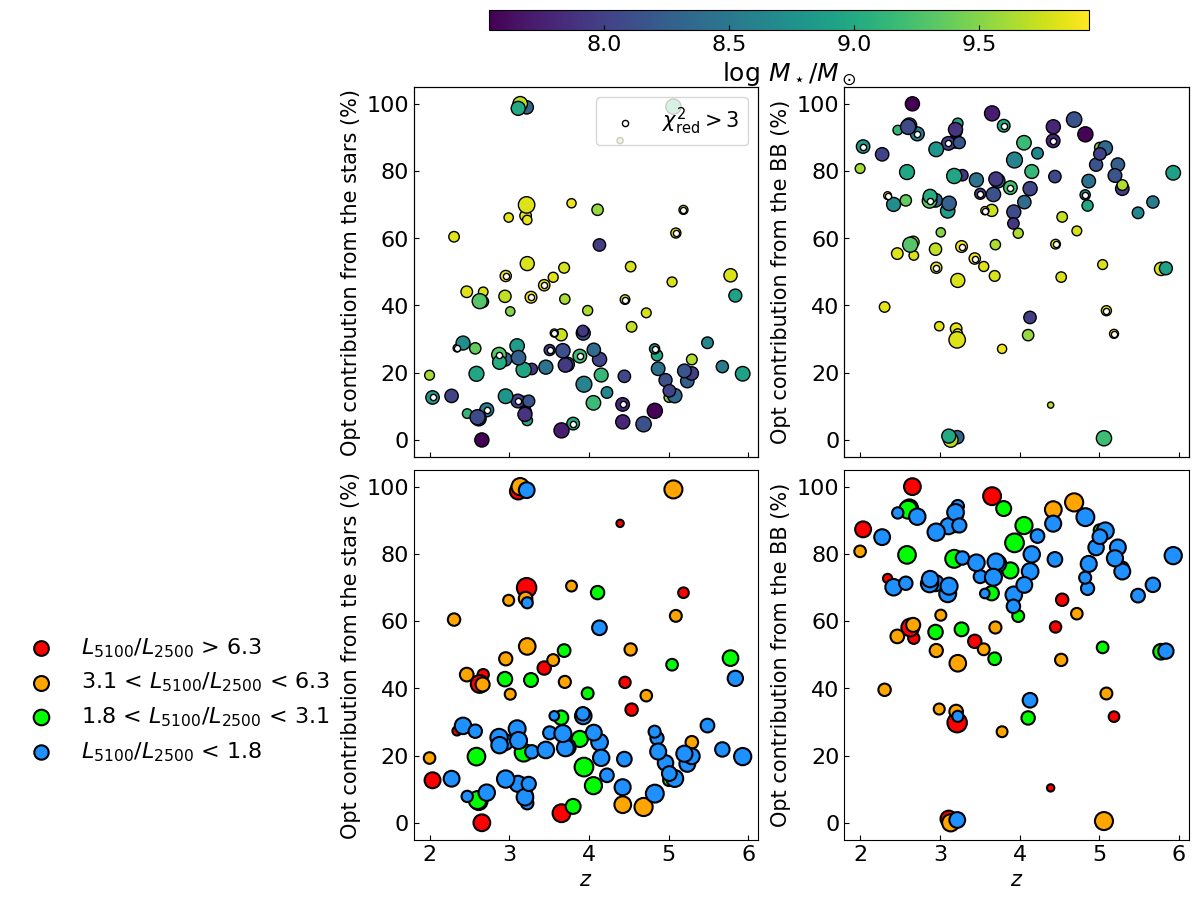}
    \end{subfigure}
    \begin{subfigure}{.4\textwidth}\includegraphics[width=\linewidth]{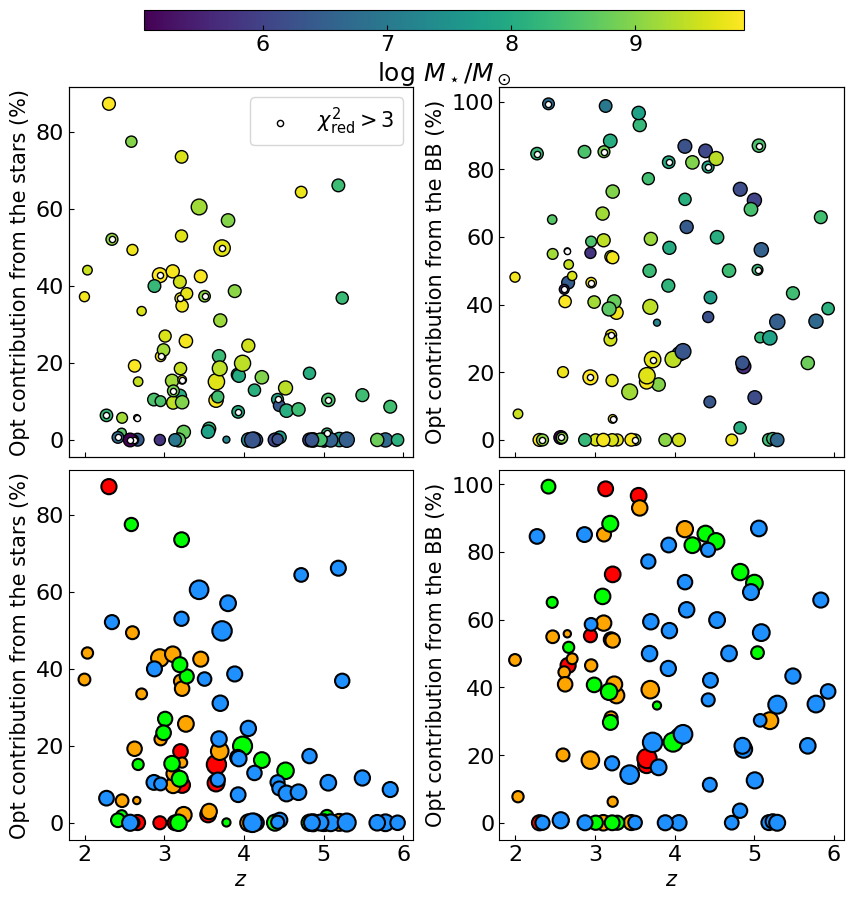}
    \end{subfigure}
    
    \caption{Evolution with $z$ of the stellar and BB contribution to the optical flux as derived following the AGN-suppressed (first panel) and AGN-agnostic (second panel) methods, respectively. In both panels, the first row shows the evolution of the components color-coded by $M_\star$. Those objects whose fits have $\chi^2_{\mathrm{red}}>3$ are highlighted with a white circle. The second row distinguishes between the 4 LRD subtypes as defined in \pg\, based on the optical-to-UV luminosity ratio. 
    The size of the markers in all the panels is defined by the temperature of the BB, with smaller markers corresponding to colder temperatures. }
    \label{fig:evolution}
\end{figure*}

Figure~\ref{fig:free_vs_suppressed} compares the $M_\star$, $A_V$, and BB temperatures obtained using the AGN-agnostic and AGN-suppressed methods.
Most objects that required high $M_\star$ values in the AGN-agnostic scenario are still massive after suppressing the AGN continuum. LRDs compatible with being BH$^*$-hostless (including some Hybrid objects and a Type I AGN) are assigned larger $M_\star$ when the BH$^*$ scenario is forced. 
In many cases, estimates of $A_V$ agree between methods. LRDs that exhibited high $A_V$ values in the AGN-agnostic scenario are now assigned lower values. However, there are still some instances in which $A_V > 1.5$.

In summary, suppressing the AGN does not automatically lead to the BH$^*$ scenario. Although the AGN contribution to the UV continuum is excluded by design, a significant contribution from an additional component (i.e., stellar emission) is often still required in many LRDs to reproduce the optical continuum. In some cases, large $A_V$ values are still inferred that are difficult to reconcile with the observed dust masses in LRDs (see Sect.~\ref{sec:discussion_1_1}). While some of these LRDs may simply not be compact enough under alternative LRD classifications, the persistence of a substantial Hybrid population, even in the AGN-suppressed scenario, may indicate intrinsic diversity or an evolutionary sequence within the LRD population. In the following subsection, we investigate the role of the stellar contribution to the optical continuum in more detail.

\subsubsection{A potential evolutionary sequence}

Interestingly, suppressing the AGN consistently produces similar or lower BB temperatures, particularly for objects classified as Type I AGNs and Hybrid in the AGN-agnostic scenario (see Fig.~\ref{fig:free_vs_suppressed}). Colder BB temperatures are usually accompanied by a larger stellar contribution to the optical continuum, as noted in Fig.~\ref{fig:example_classes_forced}. It remains to be explored whether this interplay between the stellar and BB contributions is a sign of evolution in LRDs or just a feature of a subpopulation of these objects. 

However, we note that the evolutionary trends explored here, particularly those based on the AGN-suppressed fits, are affected by the covariance of the model parameters, since the BB and stellar components are approximately complementary. Moreover, these solutions are statistically degenerate with the AGN-agnostic fits, and therefore should not be interpreted as direct evidence for a physical evolutionary sequence. Rather, they illustrate one possible interpretation of the data within the BH$^*$ framework. Furthermore, the incompleteness and heterogeneity of our sample make it challenging to draw strong conclusions about the evolution of the entire LRD population. Consequently, the results presented in this section should be regarded as tentative.

According to the observational work by \citet{Billand2025}, the decline in the number density of LRDs over cosmic time is driven by the acquisition of a stellar component that settles in the outskirts. As $M_\star$ increases, the characteristic ``V'' shape fades, and the
physical size of the galaxies grows. Similarly, \citet{Hainline2024arxiv} propose a trend between luminosity and optical slope, whereby lower-luminosity AGNs display bluer slopes as a consequence of the star-forming host galaxy becoming more dominant in these weak AGNs.

From a theoretical perspective, \cite{Kido2025} propose a different scenario, pointing out that a decrease in the rate of infalling mass onto the BH envelope, combined with BH mass growth, could lead to the dissolution of the gas envelope. Likewise, the theoretical work by \citet{Inayoshic} argues that supernova explosions from massive stars in nuclear starbursts can inject sufficient energy and momentum to expel gas from the nucleus, quenching the gas supply to the envelope and ultimately driving a transition to a normal AGN phase (see also \citealt{Merida2025_stingray} for an observational point of view).

A common feature of these scenarios is the impact of the stellar component, which can either increase the size of the system and make it more extended or indirectly induce the envelope dissolution. 
In Fig.~\ref{fig:evolution} we investigated whether the stellar contribution evolves with $z$ or if a strong stellar contribution is simply a feature observed in a subpopulation of LRDs regardless of $z$. We computed this contribution as the median ratio of the model component to the total best-fitting solution in the rest-frame optical.

According to the AGN-suppressed method (top row in panel a) the stellar contribution increases towards lower $z$, where the most massive galaxy hosts are also located. This increase in the stellar component is negatively correlated with the BB temperature, as shown by the increasingly smaller size of the markers towards lower $z$, indicating low BB temperatures. The opposite trend to that observed for the stellar component emerges when examining the BB component, whose contribution decreases towards lower $z$.
The observed decrease in BB temperature toward lower $z$, together with the increasing stellar contribution and $M_\star$ values, may suggest an evolutionary sequence. In this sequence, early LRDs would correspond to compact, heavily embedded systems dominated by hot reprocessed emission, while lower-$z$ systems would represent more evolved stages where stellar emission progressively emerges as the envelope expands, cools, or disperses.

These trends are more difficult to identify when the AGN-agnostic method is assumed (top row in panel b). While the stellar contribution evolves similarly to panel (a), the BB component does not exhibit a clear trend. The AGN component is also playing a role in this case, as it is not being suppressed, which makes interpretation more challenging. Despite the limitations, the possibility of an underlying evolutionary sequence remains intriguing and warrants further investigation in future work.

A useful point of comparison is provided by \pg, who also examined the contribution of stars to the optical continuum and the potential role of the host galaxy in LRD evolution. In that study, four LRD subtypes were defined based on the optical-to-UV luminosity ratio of a sample of 249 LRDs at $2.3 < z < 9.3$. As a result of their semi-empirical UV-to-optical spectral fitting (see Sect.~\ref{sec:intro}), they found that the stellar contribution to the optical continuum increases as the optical-to-UV ratio decreases. In other words, the contribution of the host galaxy to the optical continuum increases as the ``V'' shape becomes less pronounced, with systems such as \textit{The Cliff}, showing a negligible stellar contribution to the optical. 

\pg\, argue that LRDs such as \textit{The Cliff} would correspond to an early evolutionary phase, after which evolution would progress to bluer LRDs with an increasing stellar contribution. Nevertheless, they point out that discussions on evolutionary pathways based on $M_\star$ values are challenging due to the unknown contribution of stars to the rest-frame optical and near-IR. Unlike our observed trend of $M_\star$, their $M_\star$ values remain fairly constant across LRD subtypes, with $M_\star\sim10^{8.3}M_\odot$. The dust attenuation acting on the stellar component also remains fairly constant, with $A_V \sim 0.8$~mag. 
However, our method goes beyond a semi-empirical approach, enabling us to investigate the stellar contribution to the full spectrum. Therefore, the results reported in this work reinforce the idea that LRDs may evolve into more traditional AGN systems as a result of the growth of their host galaxies over cosmic time.

Figure~\ref{fig:evolution} (bottom rows in each panel)  also shows the redshift evolution of the stellar and BB components, separating our LRD sample into different subtypes according to the \pg\, classification. We find a trend opposite to that reported in \pg: LRDs with more extreme optical‑to‑UV ratios tend to exhibit a larger stellar contribution and are preferentially found at lower $z$. \textit{The Cliff} constitutes a notable exception, with $L_{5100}/L_{2500}>6.3$ yet a negligible stellar component. 

Our results suggest that LRDs with more pronounced ``V'' shapes may correspond to later evolutionary stages, while systems with stronger UV emission may represent earlier phases in the sequence.
Nevertheless, and as noted previously in this section, this interpretation should be treated with caution. These results are based on the analysis of the continuum, lacking the information encoded in the emission lines.
Moreover, the inferred solutions (i.e., AGN-agnostic vs AGN-suppressed) remain highly degenerate: in most cases, an AGN origin for the UV emission cannot be ruled out.
These findings may therefore indicate the presence of distinct subpopulations within the broader LRD class, superimposed on any putative evolutionary sequence, or may reflect intrinsic limitations of the BH$^*$ framework itself. 

Such limitations are highlighted by the presence of high-ionization lines in some LRDs (e.g., \citealt{Labbe2024environment}, \citealt{Tang2025}, \citealt{Treiber2025}), line ratios compatible with ionization due to an AGN \citep{Sok2026}, the preference for clumpy and non-spherical configurations in some cases (e.g., \citealt{Hviding2026}, \citealt{Sok2026}, \citealt{Sneppen2026b}), and a potential contribution of the AGN to the UV continuum (e.g., \citealt{Sun2026}, \citealt{Zhuang2025}).
These considerations highlight the need to explore additional models in a UV‑to‑optical SED fitting framework, such as the Super‑Eddington Unification Model \citep{Madau2026}, direct‑collapse black holes (e.g. \citealt{Cenci2025}, \citealt{Baggen2026}, \citealt{Pacucci2026}, \citealt{Jeon2026}), or young globular clusters in formation \citep{Chisholm2026}.

\section{Conclusions}
\label{sec:conclusions}

In this work, we performed a consistent UV-to-optical continuum fitting of 99 LRDs located at $2<z<6$ using \textit{JWST}/NIRSpec PRISM spectroscopy from the DAWN JWST Archive. Following the method showcased in \citet{Merida2026}, we employed a modified version of \texttt{Bagpipes} that includes BB emission, subject to gas absorption, to test the performance of the BH$^*$ model in describing a sample of LRDs.

When the contribution of the AGN is left free in the code (i.e., using broad priors), only $\sim4$\% of the LRDs with statistically robust fits match the BH$^*$ model's expectations. Furthermore, none of these LRDs corresponds to a dust-free scenario, requiring $\sim0.5$~mag of dust attenuation. 
$\sim9$\% of the LRDs match quasi-BH$^*$ configuration. These are systems with BB-dominated optical continua and UV emission driven by AGN + stars (AGN leakage) or totally powered by the AGN (hostless systems). The remaining 87\% of LRDs correspond to systems in which the BB does not dominate in the optical, which is, in turn, mainly driven by stars and/or AGN emission.

Forcing a BH$^*$-like solution by manually suppressing the AGN continuum elevates the number of BH$^*$ systems to $\sim35$\%. However, these solutions are statistically indistinguishable from those obtained with an AGN-agnostic setup, which are more prone to deviating from the BH$^*$ scenario. 
Breaking this degeneracy requires additional observables, such as IR-to-submillimeter constraints, exploring the UV morphology to search for host signatures, and incorporating the information encoded in the emission lines.

On the other hand, forcing BH$^*$-like solutions still yields a large fraction of fits in which the optical continuum is dominated by stars rather than by BB emission. While these forced BH$^*$ fits allow us to explore a potential evolutionary sequence for our sample of LRDs, the presence of a significant stellar contribution to the optical continuum may instead point to intrinsic limitations of the BH$^*$ model. This interpretation is further supported by independent evidence, such as the detection of high‑ionization emission lines in some LRDs and the preference for non‑spherical or clumpy geometries in previous LRD studies. These considerations suggest that additional physical scenarios should be evaluated alongside the BH$^*$ model within a UV-to-optical SED-fitting framework.

\textit{Data availability}. Mérida, R. M., Sawicki, M., Willott, C., Gaspar, G., \& Iyer, K. (2026). Appendices from UV-to-optical insights into the BH$^*$ model in little red dots. Zenodo. https://doi.org/10.5281/zenodo.22057251

\begin{acknowledgements}
This research was enabled by Canadian Space Agency grants 18JWST-GTO1 and 24JWGO3A12 and Natural Sciences and Engineering Research Council (NSERC) of Canada grants RGPIN-2020-06023, RGPAS-2020-00065, and RGPIN-2026-08047.
This research used the Canadian Advanced Network For Astronomy Research (CANFAR) platform operated in partnership by the Canadian Astronomy Data Centre and The Digital Research Alliance of Canada with support from the National Research Council of Canada, the
Canadian Space Agency, CANARIE, and the Canada Foundation for Innovation.\\
This work is based on observations made with the NASA/ESA/CSA James Webb Space Telescope. The data were obtained from the Mikulski Archive for Space Telescopes at the Space Telescope Science Institute, which is operated by the Association of Universities for Research in Astronomy, Inc., under NASA contract NAS 5-03127 for JWST. 
These observations are associated with programs ERS-1345; GTO-1180, 1181, 1210, 1286, GO-1895, 1963; GO-1837; GTO-1208, GO-3362; GO-2561; GO-5105; GO-2514; GO-4233; GTO-1211; GO-6398; GO-5224; GO-2198; DD-6585; DD-2750; and GO-4106.\\
The data products presented herein were retrieved from the Dawn JWST Archive (DJA). DJA is an initiative of the Cosmic Dawn Center (DAWN), which is funded by the Danish National Research Foundation under grant DNRF140.
\end{acknowledgements}

\bibliographystyle{aa}
\bibliography{aa61390-26}

%

\end{document}